# A salt water battery with high stability and charging rates made from solution processed conjugated polymers with polar side chains

Davide Moia,*,1,‡ Alexander Giovannitti,*,1,2,‡ Anna A. Szumska,1 Martin Schnurr,2 Elham Rezasoltani,1 Iuliana P. Maria,2 Piers R.F. Barnes,1 Iain McCulloch,2,3 Jenny Nelson*,1

1 Department of Physics, Imperial College London SW7 2AZ London, UK

2 Department of Chemistry, Imperial College London SW7 2AZ London, UK

3 Physical Sciences and Engineering Division, KAUST Solar Center (KSC), King Abdullah University of Science and Technology (KAUST), KSC Thuwal 23955-6900, Saudi Arabia

* davide.moia11@imperial.ac.uk; a.giovannitti13@imperial.ac.uk; jenny.nelson@imperial.ac.uk

‡ These authors contributed equally to this work

**Abstract**

We report a neutral salt water based battery which uses p-type and n-type solution processed polymer films as the cathode and the anode of the cell. The specific capacity of the electrodes (approximately 30 mAh cm$^{-3}$) is achieved via formation of bipolarons in both the p-type and n-type polymers. By engineering ethylene glycol and zwitterion based side chains attached to the polymer backbone we facilitate rapid ion transport through the non-porous polymer films. This, combined with efficient transport of electronic charge via the conjugated polymer backbones, allowed the films to maintain constant capacity at high charge and discharge rates (>1000 C-rate). The electrodes also show good stability during electrochemical cycling (less than 30% decrease in capacity over >1000 cycles) and an output voltage up to 1.4 V. The performance of these semiconducting polymers with polar side-chains demonstrates the potential of this material class for fast-charging, water based electrochemical energy storage devices.

**Introduction**

The development of inexpensive, robust high performance electrochemical storage devices is expected to be enabled by the development of new electrode and electrolyte materials.[1,2,3] Expansion of predominant battery technologies – such as lithium ion devices – is limited by concerns over toxicity and cost of electrodes and, when used with organic electrolytes, safety.[4,5] Viable alternative electrode materials must meet certain performance criteria in terms of energy and power density. These include: (i) Efficient electronic and ionic transport to enable the full electrode volume to be accessed rapidly during electrochemical charging and discharging; (ii) oxidation (cathode) and reduction (anode) potentials that approach the boundaries of the electrochemical window of the electrolyte used, in order to maximize operational voltage; and (iii) a high density and individual capacity (or range of oxidation state) of redox sites. The criteria should, ideally, be met with materials systems that offer low cost, high availability and low toxicity. For operational durability the electrode should also be robust to electrochemical stress and insoluble in the working electrolyte.[6]

Organic electrode materials are a promising option[7,8,9] on account of the tuneability of their properties and the potential to be processed easily into different forms, which include porous nanostructures or nanocomposites that facilitate ion penetration.[10,11,12] Conjugated molecular materials are especially interesting because they offer a wide range of oxidation and reduction potentials, tuneable via chemical structure, and because π-conjugation facilitates electronic injection and transport. They can be designed to operate in water-based rather than organic electrolytes, bringing advantages in terms of toxicity and safety,[13,14,15,16] although finding stable conjugated polymers with the very negative reduction potentials needed to maximize operational voltage in water remains a challenge.[17,9] One perceived limitation of organic electrodes is their volumetric capacity which is controlled by the density of individual redox sites and their range of oxidation state which limits the achievable charge density under charging; this may be further compromised by the use of non-redox-active side chains for polymer processability and ionic access as well as by blending the polymer with additional conductive or insulating scaffolds.[18,19]

Approaches to electrochemical energy storage using conjugated molecular materials have included: the use of conjugated polymers to conduct charges to and from redox-active units,[20] conjugated polymers with pendant redox-active groups[21,22] and non-conjugated polymers bearing either in-chain[14,23] or pendant redox-active groups,[22] including in some cases radical groups that are easily charged.[15,24] Encouraging specific capacities of hundreds of mAh g$^{-1}$ have been reported[14] for some systems along with high cycle-life but with limited charging and discharging rates. Polymer chain conjugation has been shown to enhance charging rates, apparently by facilitating electronic transport,[18] but only to around 100 C. Most reported systems operate in organic or acidic electrolytes and whilst some work with aqueous salt electrolytes they tend to require high salt concentrations. A primary challenge, therefore, appears to be achieving high ionic and electronic conductivity simultaneously, to enable the electrodes to charge and discharge efficiently in aqueous electrolytes.

Here, we present a design strategy for conjugated polymer electrodes whereby we exploit side chain design to enhance ionic conductivity and separately optimize backbone and side-chain structures to enable mixed electronic-ionic conduction. First, we identify conjugated polymers with a low ionization potential, here referred to as p-type, or low reduction potential, here referred to as n-type, to address specifications for both cathode and anode materials with redox potentials which complement the allowed electrochemical window for water (in this context, 'p-type' and 'n-type' simply indicate their relative energetic 'ease' for oxidation and reduction, respectively, and not extrinsic doping). Secondly, we design polymer side chains based on ethylene glycol based structures (glycol) for the p-type and zwitterion structures for the n-type polymer. These side chains were chosen to favor conduction of alkali metal cations and halide anions, and to enable the use of neutral pH, sodium chloride water based electrolyte. A similar strategy was previously applied to the design of polymers for organic electrochemical transistors for bioelectronics, resulting in stable p- and n-type polymers with specific capacitances >350 F cm$^{-3}$ using NaCl:deionized water (DIW) as electrolyte.[25] The n-type polymer structures presented here illustrate the potential to further improve the specific capacity and application potential of these materials through side chain engineering. Charging and discharging of ~100 nm thick polymer films occurs efficiently on the second timescale. Our material design strategy addresses the three factors described above in that it enables efficient ion transport within the polymer films for fast charging/discharging; it results in p-type and n-type polymers with redox activity which takes advantage of nearly the full available electrochemical window of water; and it results in highly reversible electrochemical charging which is extended to the whole polymer film. We demonstrate battery devices combining the p- and n-type conjugated polymers as the cathode and the anode of a water based electrochemical storage cell that can be operated under a nearly unipolar potential window of 1.4 V at >1000 C-rate (i.e. using current values which charge/discharge the battery in 1/1000 hours). The device is a demonstration of a solution processable polymer battery working under neutral pH water conditions and showing rate capabilities comparable to supercapacitors.

**Polymer materials for water based batteries**

Our design strategy for both p-type (cathode) and n-type (anode) materials is to attach polar side chains to semiconducting backbone structures (Figure 1). The p-type, a homo-alkoxybithiophene polymer p(gT2) (Figure 1a) presents an electron rich structure with bithiophene units planarized relative to unsubstituted bithiophene by the non-covalent sulfur oxygen interactions involving the oxygen atoms in the 3,3' position of the bithiophene.[26] Methyl end-capped triethylene glycol side chains (TEG) attached to each thiophene yield high polymer solubility in *N,N*-dimethylformamide (DMF) and chloroform. In addition, glycol side chains are known to assist anion transport effectively[27,28] and expected to enhance charging and discharging rates in aqueous solution. The n-type is a donor-acceptor type copolymer based on naphthalene-1,4,5,8-tetracarboxylic diimide (NDI) and alkoxybithiophene (Figure 1b).[25,29] This structure is known to exhibit a low electrochemical reduction potential.

We attached zwitterion side chains at the NDI acceptor unit in order to control interactions between side chain and positive ions during electrochemical charging. NDI copolymers with pendant zwitterions have been previously applied as interlayers to improve cathode selectivity in organic solar cells.[30,31] Here, for the attachment of the polar zwitterion side chains on the NDI unit, we developed a synthesis involving post-functionalization reactions of a conjugated polymer containing dimethylamino functional groups. First, we prepared a copolymer with dimethylamino functional groups. Then, the dimethylamino groups were converted into ammonium bromides featuring an ester group at the end of the side chain. Finally the ester was cleaved to form the zwitterion side chain copolymer p(ZI-NDI-gT2). The glycol side chains on the alkoxybithiophene unit were attached to ensure high solubility in organic solvents. This design strategy yielded a copolymer which is soluble in polar organic solvents including dimethyl sulphoxide and methanol but insoluble in water, an important criterion for the development of redox-active materials in aqueous electrolytes. Whilst zwitterion functionalised NDI polymers have been reported before[30] our strategy allowed the preparation of the zwitterion copolymer to be done in solution, enabling diversity in functional groups and also avoiding the use of toxic solvents. Details of the co-polymer synthesis, characterization and properties are presented in Supplementary Sections 1-7.

**Results and discussion**

Cyclic voltammetry (CV) measurements of both p-type (p(gT2), Figure 2a) and n-type (p(ZI-NDI-gT2), Figure 2b) polymers deposited on fluorine doped tin oxide (FTO) conducting glass substrates show a current profile with large area, when measured in a three electrode electrochemical cell with 0.1 M NaCl:DIW used as supporting electrolyte. The large area enclosed by the CV is beneficial for the function of charge storage.[32] The oxidation (reduction) processes are attributed to charging of the film via injection of holes (electrons) from the FTO substrate into the polymer and ionic charge compensation *via* exchange of ions between electrolyte and polymer film. The scans show high reversibility and stable peak position after the first cycle (Further experimental details are given in Supplementary Section 8).

Spectroelectrochemical measurements (Figure 2c and d) show how optical spectral evolution follows the electrochemical charging of both polymers. For the p-type polymer the main absorption feature in the visible (peak at 645 nm) is completely quenched upon oxidation up to 0.2 V and a new band appears in the NIR; we attribute this to the bleach of polymer neutral state absorption, and formation of positive polarons. Further oxidation ($V > 0.2$ V vs Ag/AgCl) reduces the NIR absorption suggesting that further conversion occurs, which we ascribe to bipolaron formation. Calculated spectra of the trimer (gT2)$_3$ in its neutral, singly and doubly charged state using TD-DFT shown in Figure 2e (further details are presented in Supplementary Section 9) support our assignment of the spectral changes to polaron and bipolaron formation, as does a prior report of polythiophene spectroelectrochemistry.[33] Importantly, the spectral evolution suggests complete conversion of neutral polymers into the charged state, indicating that the full volume of the electrode undergoes charging. Scanning to more positive potential shows that higher levels of charging can be achieved, although this compromises the coulombic efficiency of the electrode (Supplementary Section 10).

For the n-type polymer the spectroelectrochemical data suggest reduction of the polymer chains resulting in the formation of an electron polaron at a potential of -0.4 V and bipolaron between -0.4 V and -0.75 V vs Ag/AgCl. Both these processes are reversible over multiple cycles. Our assignment of the spectral features to polaron and bipolaron formation are again supported by calculated absorbance spectra (normalized) of neutral and charged monomers (Figure 2f).

Cyclic voltammetry measurements on p-type and n-type polymer films of different thicknesses at a fixed scan rate (50 mV s$^{-1}$) allowed us to estimate the specific capacity of the films at 25 mAh cm$^{-3}$ for the p(gT2) and 36 mAh cm$^{-3}$ for the p(ZI-NDI-gT2) polymers; these represent charge loadings of around 0.6 holes per alkoxybithiophene repeat unit (gT2) and 1.4 electrons per ZI-NDI-gT2 repeat unit, which approach the theoretical limit of these materials (see Supplementary Section 11 and 12). Continuous cycling showed that 70% of the initial (2$^{nd}$ scan) capacity is retained after 1000 cycles in water for both polymers (Supplementary Section 13).

*Water based polymer battery*

The encouraging specific capacity, charging rate and stability of the n- and p-type polymer electrodes suggests that they may perform well in a two-electrode battery structure. We construct such an electrochemical charging device using the n-type p(ZI-NDI-gT2) polymer film characterized in Figure 2b and d and a p-type p(gT2) film with similar capacity as anode and cathode in an electrochemical cell filled with a salt-water (0.1M NaCl:DIW) electrolyte. The expected operation of the device is illustrated in Figure 3a. At equilibrium, the electrochemical potential of anode, electrolyte and cathode is constant across the device (black dashed line in Figure 3c). Note that the precise value of this potential depends on the state of charge of each electrode and may vary over lifetime due to imbalanced charge retention performance in the two electrodes (see discussion below). When a positive potential $V_{battery}$ is applied to the cathode with respect to the anode, Na+ ions are attracted to the anode where, as we discuss below, they become electrostatically bound to the negatively charged polymer while Cl- ions are attracted to the cathode and bind to the positively charged p-type polymer. The absolute values of the half-cell potentials $V_n$ and $V_p$ (where $V_{battery} = V_p - V_n$) are not known but their values should not extend beyond the potentials for reduction and oxidation of water (indicated in Figure 3b) for stable operation in water.

The measured CV response of the two-electrode device encloses a large area, which is consistent with the CVs of the individual electrodes as illustrated in Figure 3c. The device can be operated at an applied potential of 1.4 V which is the sum of potential ranges used to characterize the p-type (0.5 V vs Ag/AgCl) and n-type (-0.9 V vs Ag/AgCl) polymers. Figure 4a and 4c show the spectroelectrochemical characterization of the polymer battery for a cyclic voltammetry measurement run at 100 mV s$^{-1}$ (black line in Figure 4a). The spectroelectrochemical response for the two-electrode device (Figure 4c) shows, by comparison with individual electrode spectra, that electrochemical charging of the battery device produces the bipolaron in both films. Notably, integration of the reversible current from the CV data in Figure 4a shows that 78% to 94% (depending on rate of discharge) of the total maximum stored charge can be extracted at positive voltages, while galvanostatic measurements suggest that this value is over 90% for the range of C-rate considered here. These results indicate that the device effectively implements the function of energy storage.

The battery device exhibits a specific capacity above 15 mAh cm$^{-3}$ at C-rates >1000. Note that because the battery's specific capacity accounts for both electrodes it is about half the value for the individual polymers. The fast charging/discharging capability is confirmed by the relative insensitivity of the CV to scan rate up to 200 mV s$^{-1}$ and the linear dependence of peak current on scan rate up to 2000 mV s$^{-1}$ (Supplementary Section 14). The peak shift in the cyclic voltammograms and the drop in capacity at fast scan or charging rate (Figure 4b) are consistent with the presence of a series resistance due to the two FTO layers in the order of 100Ω. Further evidence that response is rate limited by series resistance and internal charge-transfer resistance is provided (Supplementary Sections 15-17). Finally a stability measurement, displayed in Figure 4d, was run on the cell, showing less than 20% drop in capacity over 1000 cycles.

*On the mechanism of polymer charging and discharging*

The polymer thin films investigated in this work can be efficiently charged and discharged at fast rates. We attribute the high rate capability of these electrodes to the structure of the polymer side chains which is suitable for the transport of ions within the film. We found that glycol side chains are a favorable medium for the transport of chloride ions in the case of the p(gT2) polymer. We assign this to the limited ability of the glycol chains to interact with anions. Oxygen atoms in this structure have a strong tendency to chelate positive ions.[34,35,26] Carbon atoms on the other hand, despite being polarized, are not able to chelate negative ions due to the unavailability of unpaired electrons (sp$^3$ hybridization) and as a result they enable and do not restrict the transport of anions. For example, fast switching times can be observed when polymer semiconductors including glycol side chains are operated in organic electrochemical transistors (OECTs) where anions need to migrate into the bulk of the polymer.[27,28]

Regarding the (zwitterion) p(ZI-NDI-gT2) polymer, DFT calculations (see Supplementary Section 9 for full discussion) show that injected negative charges are expected to localize on the NDI unit, which suggests that the side chain on that unit could help to control electronic-ionic charge interactions. We hypothesize that cations do not need to migrate close to the backbone of p(ZI-NDI-gT2) upon charging of the electrode due to the presence of a permanent positive charge on the ammonium group close to the NDI unit. When the polymer is in its neutral state, the positive charge on the ammonium group is compensated by the carboxylate ion of the zwitterion side chain. When the polymer is reduced, the ammonium ion can compensate the negative charge on the backbone, which we expect to localize on the NDI unit, leaving the negative carboxylate group uncompensated. This would encourage sodium ions to interact with the carboxylate group of the zwitterion side chain rather than to approach the polymer backbone, facilitating the attachment and release of ions upon charging. A schematic of the proposed mechanism is shown in Figure 1c. In order to explore this hypothesis we synthesized and characterized another n-type polymer, p(g7NDI-gT2), where glycolated chains are attached to the NDI unit instead of the zwitterion chain. Glycol chains allow penetration of sodium ions into the polymer film but may be expected to inhibit cation transport due to the chelation by oxygen atoms.[36] Accordingly, we observed an upper limit to reversible charging of p(g7NDI-gT2) when scanning to potentials beyond the first reduction peak. A similar behavior was also reported previously for OECTs using polymers with glycol chains.[25] We tentatively assign the larger reversible specific capacity of the (zwitterion) p(ZI-NDI-gT2) polymer, compared with the case of the (glycolated) p(g7NDI-gT2) polymer, to the different nature of the interaction between sodium ions and electron polarons which appears to allow the bipolaron to form. The characterization of the p(g7NDI-gT2) polymer is presented in Supplementary Section 18.

Improvements in ionic conductivity through the use of glycol and zwitterion groups have previously been presented for block copolymer structures and for non-conjugated radical polymers.[37,6,15,38,39,40,41,42] We assign the improved charging rates for our polymers relative to most other varieties of polymer electrodes to the higher expected charge carrier[28] and ionic[27] mobilities, enabled by the polar (glycol and zwitterion) side chains and extended backbone conjugation. Whilst the specific capacity achieved in our materials is lower than in the best organic electrodes, it approaches the limit defined by the redox site density, and can be improved by careful optimization of side chain length and structure. A further important target for material and device optimization is to improve and balance retention of charge in both electrodes, currently limited to thousands of seconds as we show in Supplementary Section 19. In the same Section we discuss how charge retention issues affect device cyclablility, showing an example of capacity recovery after continuous cycling.

**Conclusions**

We have presented a strategy for material design that enables the fabrication of solution processable p-type and n-type polymers that can be employed as cathode and anode of water based battery devices. The use of polar side chains is identified as a successful solution to improve the ion penetration within the conjugated polymer films. We show that sacrificing volume density of redox sites to include a more favorable medium for ion transport within the battery electrode can be a 'good investment' to achieve fast rates of reversible charging/discharging of the polymer chains. Specific capacity of the electrodes in the order of 30 mAh cm$^{-3}$ and <30% drop in capacity over >1000 cycles were observed. We show that the process of charging and discharging occurs on the second timescale (C-rate > 1000) for both p(gT2) and p(ZI-NDI-gT2) polymers with thickness up to about 100 nm, expected to be limited by contact series resistance. These p-type and n-type polymers were coupled in a two terminal structure with a neutral NaCl aqueous supporting electrolyte to form a battery with > 15 mAh cm$^{-3}$ specific capacity, which was measured up to > 2000 C-rate. The design of the polymers' backbone structure resulted in favorable energetics and ultimately in operational voltage of the battery up to 1.4V. This study shows that conjugated polymers with polar side chains can be used as electrodes of battery devices where these materials implement the transport of both the electronic and the ionic charges within their bulk as well as the redox behavior, enabling high levels of power density. These polymers are also solution processable, and therefore compatible with printing techniques and roll to roll high throughput fabrication. In addition, the use of sodium chloride and water at neutral pH values as electrolyte is a promising demonstration for a potentially inexpensive and safe electrochemical energy storage solution.

**Acknowledgements**

We thank Peter R Haycock for the fruitful discussions about the NMR spectra. DM, PB and JN are grateful for funding from the EPSRC Supersolar Hub (grant EP/P02484X/1). This project has received funding from the European Research Council (ERC) under the European Union's Horizon 2020 research and innovation program (grant agreement No 742708), IMEC Synergy Grant SC2 (610115), EPSRC Project EP/G037515/1, EP/M005143/1, EP/N509486/1 and from The Imperial College Faculty of Natural Sciences Strategic Research Fund.

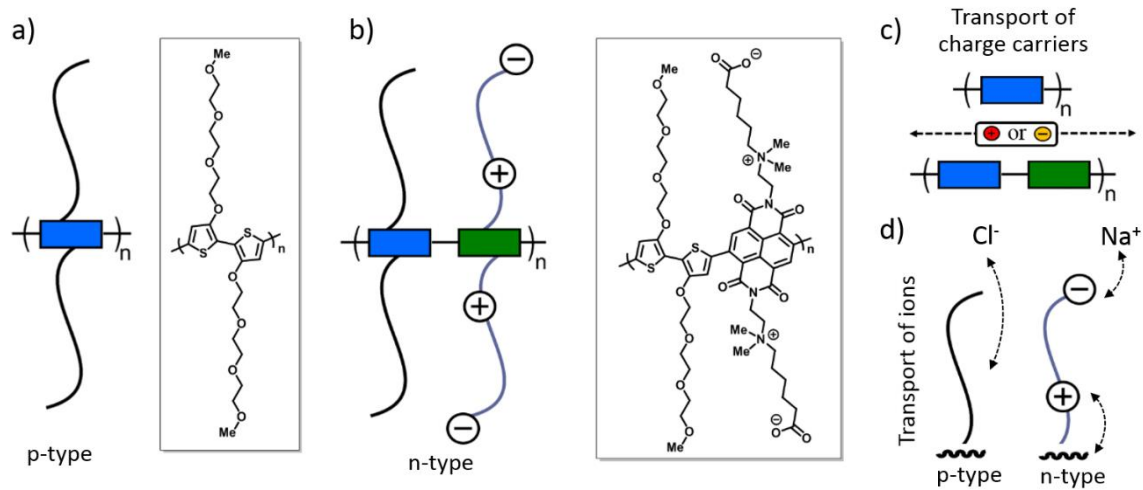

Figure 1. Simplified structures and molecular structure of the polymers which are used in this work: (a) p-type homo-polymer with glycol side chains; (b) n-type donor-acceptor copolymer with zwitterion side chains on the acceptor units and glycol side chains on the donor units. (c) Transport of the electronic charges along the backbone of the polymers and (d) transport of ions with the aid of polar side chains.

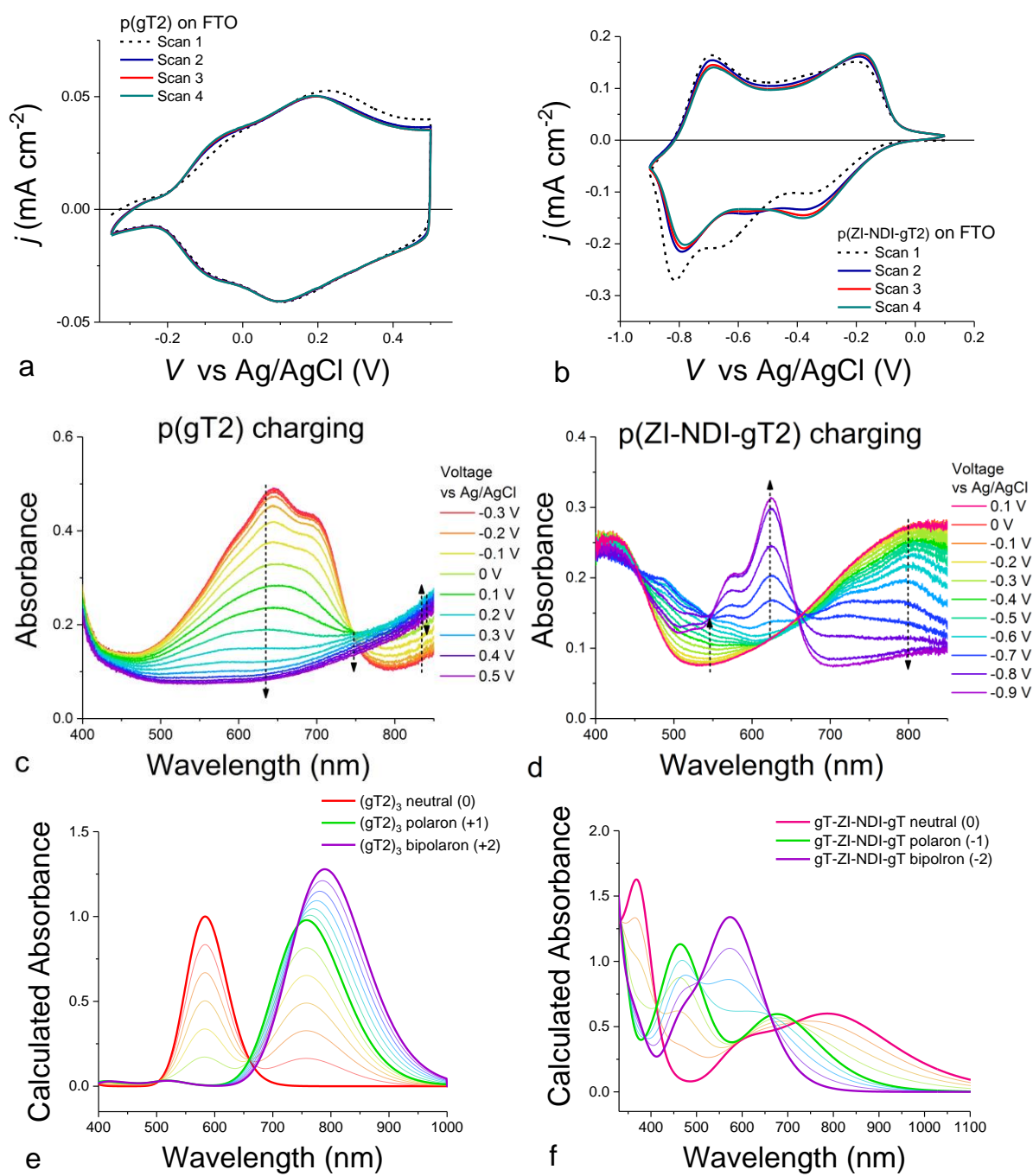

Figure 2. Spectroelectrochemistry of (left) p(gT2) and (right) p(ZI-NDI-gT2) polymers on FTO in 0.1M NaCl:DIW electrolyte. (a, b) First 4 cyclic voltammetry scans performed on the samples using scan rate of 50 mV s$^{-1}$. Before the measurement shown in (b) the electrolyte was degassed for 15 minutes with argon gas to remove molecular oxygen. (c, d) UV vis absorbance spectra for the charging of the polymer films during the first cyclic voltammetry scan as a function of potential applied to the sample vs Ag/AgCl. (e, f) Normalized absorbance calculated with TD-DFT for the trimer (gT2)$_3$ and the monomer gT-ZI-NDI-gT in the neutral and charged (polaron and bipolaron) states. Linear combinations of these three spectra were used to plot spectra of intermediate states.

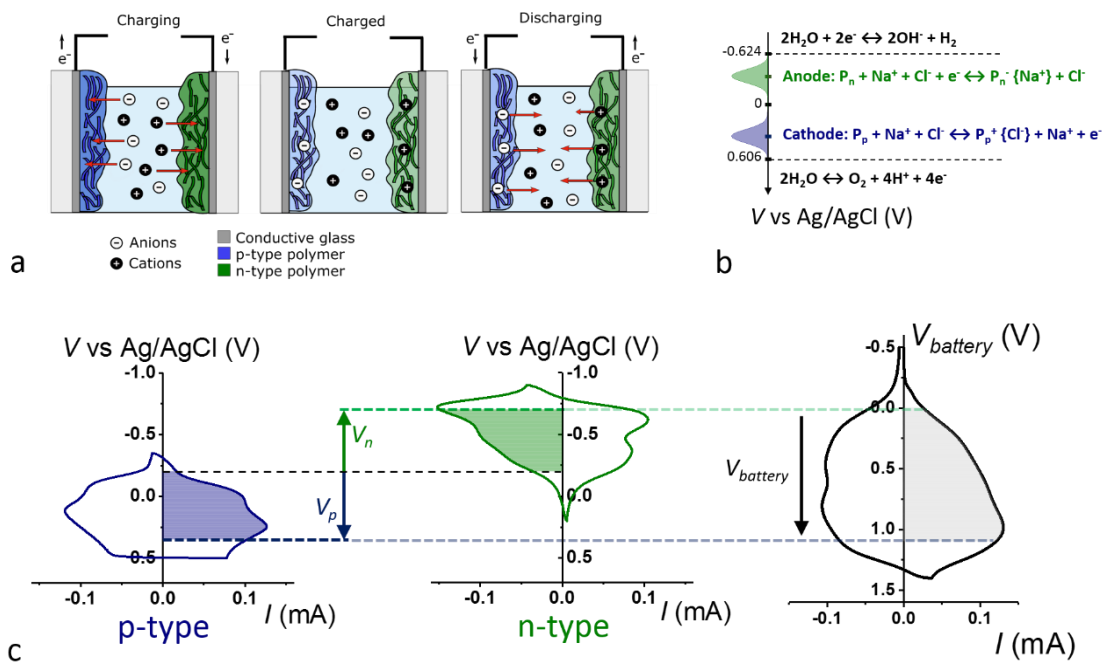

Figure 3. (a) Schematics of the working mechanism of a water based polymer battery. (b) Reactions at the cathode and at the anode and electrochemical window of water at neutral pH. (c) Energetics and charge distribution in a polymer battery: (left and center) cyclic voltammetry measurements on the p-type and the n-type polymers in a three electrode cell at 50 mV s$^{-1}$ (potentials are referenced to Ag/AgCl) and (right) cyclic voltammetry measurement on the complete battery at a scan rate of 100 mV s$^{-1}$. The value $V_{battery}$ is the difference in electrochemical potentials between the p-type (cathode) and n-type (anode) polymers.

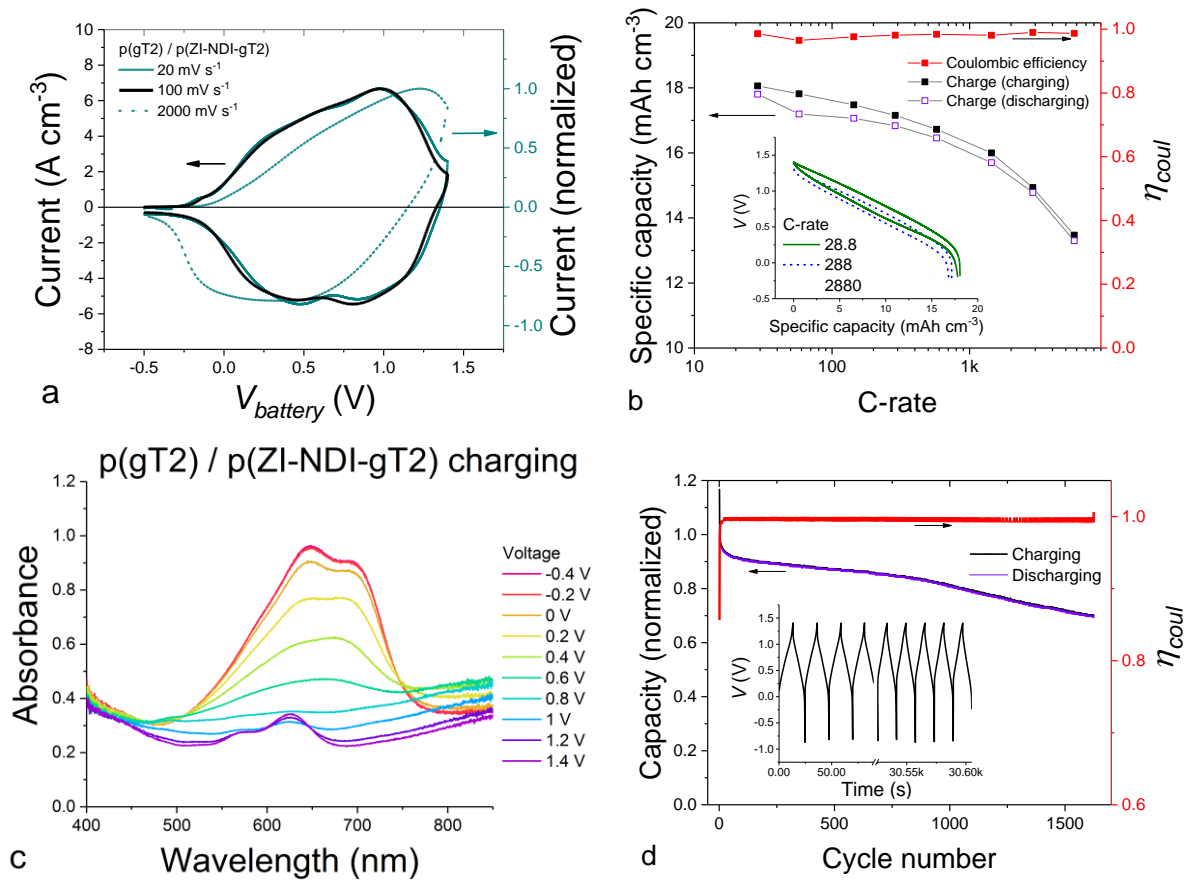

Figure 4. Polymer battery with structure FTO / p(gT2) (87 nm)/ 0.1M NaCl:DIW / p(ZI-NDI-gT2) (70 nm) / FTO. Voltage is applied/measured at the cathode (p(gT2) electrode) with respect to the anode (p(ZI-NDI-gT2) electrode). (a) Cyclic voltammetry measurements performed at different scan rates. (b) Specific capacity and coulombic efficiency as a function of C-rate (inset shows the galvanostatic charge-discharge curves at different C-rates). (c) Evolution in optical absorbance for the two films in series corresponding to the charging of the cell at 100 mV s$^{-1}$. (d) Normalized specific capacity of the battery and coulombic efficiency measured as a function of cycle number. The measurement was performed using galvanostatic cycling at approximately 300 C-rate (inset showing the first and last cycles of the experiment).

# Supplementary information

**A salt water battery with high stability and charging rates made from solution processed conjugated polymers with polar side chains**

Davide Moia,*,[1,‡] Alexander Giovannitti,*,[1,2,‡] Anna A. Szumska,[1] Martin Schnurr,[2] Elham Rezasoltani,[1] Iuliana P. Maria,[2] Piers R.F. Barnes,[1] Iain McCulloch,[2,3] Jenny Nelson*,[1]

[1] Department of Physics, Imperial College London SW7 2AZ London, UK

[2] Department of Chemistry, Imperial College London SW7 2AZ London, UK

[3] Physical Sciences and Engineering Division, KAUST Solar Center (KSC), King Abdullah University of Science and Technology (KAUST), KSC Thuwal 23955-6900, Saudi Arabia

[‡] These authors contributed equally to this work

* davide.moia11@imperial.ac.uk; a.giovannitti13@imperial.ac.uk; jenny.nelson@imperial.ac.uk

# Contents

# 1. Material synthesis and characterization

Column chromatography with silica gel from VWR Scientific was used for flash chromatography. Microwave experiments were carried out in a Biotage Initiator V 2.3. $^1$H and $^{13}$C NMR spectra were recorded on a Bruker AV-400 spectrometer at 298 K and are reported in ppm relative to TMS. UV-Vis absorption spectra were recorded on UV-1601 ($\lambda_{max}$ 1100 nm) UV-VIS Shimadzu spectrometers.

MALDI TOF spectrometry was carried out in positive reflection mode on a Micromass MALDImxTOF with trans-2-[3-(4-tert-Butylphenyl)-2-methyl-2-propenylidene]-malononitrile (DCTB) as the matrix.

Cyclic voltammograms were recorded on an Autolab PGSTAT101 with a standard three-electrode setup with ITO coated glass slides, a Pt counter electrode and a Ag/AgCl reference electrode (calibrated against ferrocene (Fc/Fc$^+$)). The measurements were either carried in an anhydrous, degassed 0.1 M tetrabutylammonium hexafluorophosphate (TBAPF$_6$) acetonitrile solution or in a 0.1 M NaCl aqueous solution as the supporting electrolyte with a scan rate of 100 mV/s.

Gel permeation chromatography (GPC) measurements were performed on an Agilent 1260 infinity system operating in DMF with 5 mM NH$_4$BF$_4$ with 2 PLgel 5 μm mixed-C columns (300×7.5mm), a PLgel 5 mm guard column (50x7.5mm) at 50 °C with a refractive index detector as well as a variable wavelength detector. The instrument was calibrated with linear narrow poly(methyl methacrylate) standards in the range of 0.6 to 47kDa.

Dialysis was carried out in a dialysis kit Thermo Scientific Slide-A-Lyzer Cassette with a molecular weight cut off of 2K. The deionised water was replaced every 6 h and the dialysis was carried out for two days.

End-capping procedure: After the reaction was cooled to room temperature, 0.1 mL of a solution made of 1.00 mg of Pd$_2$(dba)$_3$ and 0.1 mL of 2-(Tributylstannyl)thiophene in 0.5 mL of anhydrous degassed DMF was added and heated for 1 h to 100 °C, then 0.1 mL of a solution made of 1.00 mg of Pd$_2$(dba)$_3$ and 0.1 mL of 2-bromothiophene in 0.5 mL of anhydrous degassed DMF was added and heated to 100 °C.

# 2. Summary of the synthesized materials and their properties

## Synthesis of poly(TEG-alkoxybithiophene)

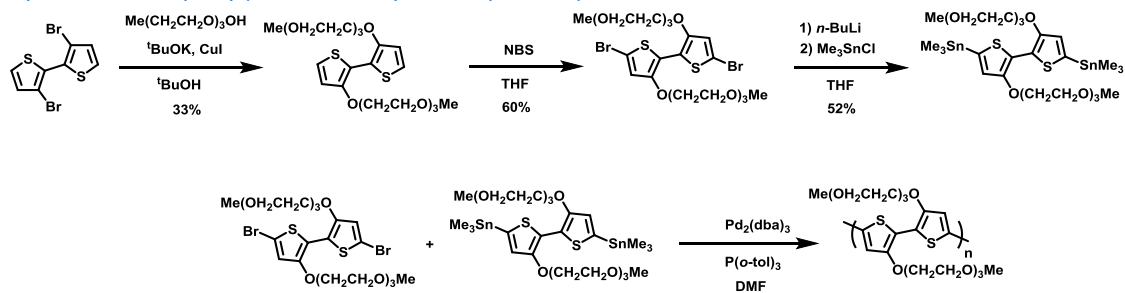

## Synthesis of p((DMA)-NDI-gT2)

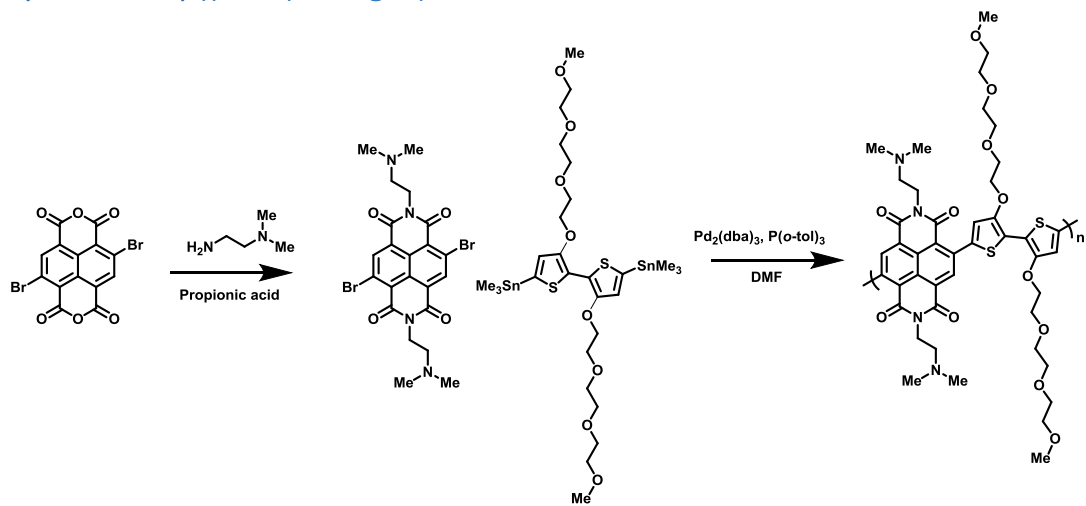

## Synthesis p(ZI-NDI-gT2)

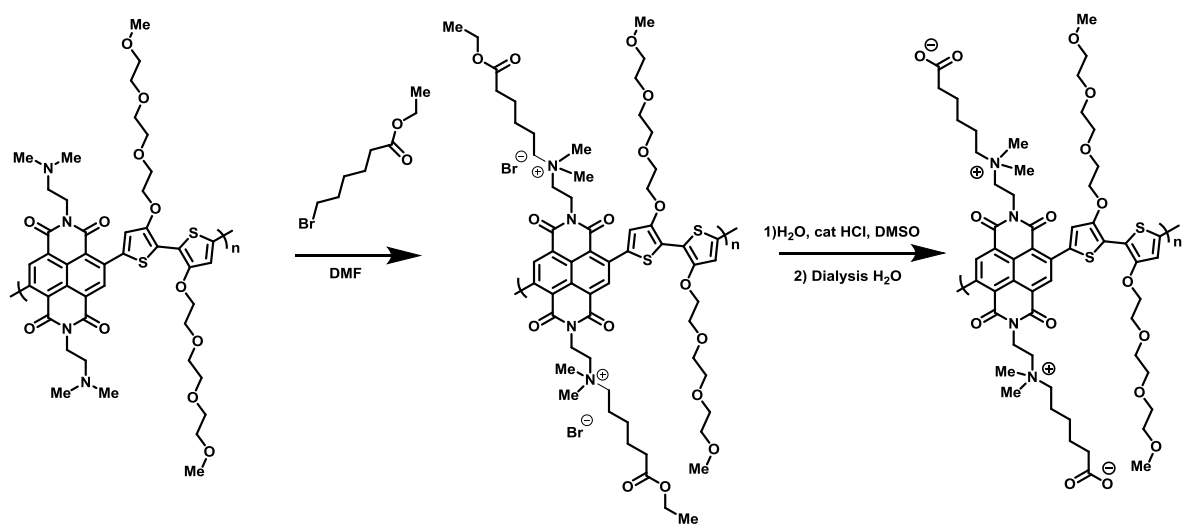

## Polymers' properties

Table 1. Summary of the polymers' properties

| Polymer | IP[A] [eV] | IP[B] [eV] | EA[B] [eV] | $M_n$[D] | $M_w$[D] |
|---|---|---|---|---|---|
| p(gT2) | 4.48 | 4.2 | - | 56 kDa | 306 kDa |
| p(ZI-NDI-gT2) | 5.15 | 5.0 | 4.0 | 24 kDa | 53 kDa |

[A] Photoelectron spectroscopy in Air (PESA)

[B] CV measurements in acetonitrile (0.1 M NBu$_4$PF$_6$, 100 mV/s)

[C] Measurements were carried out in degassed 0.1 M NaCl aqueous solution vs Ag/AgCl.

[D] GPC measurements were carried out in DMF with NBu$_4$BF$_4$.

# 3. General notes on the synthesis of the zwitterion copolymer

The zwitterion side chain on the NDI unit had to be introduced after building up the backbone since the dibromo-NDI-dimethyl amino monomer can easily undergo nucleophilic aromatic substitution. It was therefore not possible to isolate the dibromo-zwitterion NDI monomer. Several attempts to prepare the monomer resulted in the formation of oligomers where the dimethylamino group replaced the bromides on the NDI unit. This side reaction is also most likely the reason for the relatively low degree of polymerization which was observed for the copolymer containing dimethyl amino groups (p((DMA)-NDI-gT2)). The GPC results of p((DMA)-NDI-gT2) in chloroform showed a molecular weight distribution of $M_n$ = 5.3 kDa and $M_w$ = 8.9 kDa which is in agreement with results obtained by mass spectrometry.

The alkylation of p((DMA)-NDI-gT2) was carried out in DMF with ethyl-6-bromohexanoate to form the ammonium bromide copolymer p((DMA-Br)-NDI-gT2) with an ester group. To verify the selective alkylation of the dimethylamino group, nuclear magnetic resonance (NMR) spectroscopy was carried out and the formation of the ammonium bromide was observed by [1]H NMR and 2D correction spectroscopy (COSY) measurements. The ammonium bromide copolymer aggregated significantly in polar organic solvents such as DMF which is most likely the reason for the observed increase of the molecular weight distribution of p((DMA-Br)-NDI-gT2) ($M_n$ = 38 kDa, $M_w$ = 60 kDa). The ammonium bromide copolymer are water soluble and could therefore not be used in a water based energy storage device.

In the final step of the synthesis, the ester was cleaved under acidic conditions to form the carboxylate zwitterion side chain. The polymer was purified by dialysis in deionized water to remove water soluble side products such as hydrobromic acid (HBr). To verify the selective ester cleavage, [1]H NMR spectroscopy was carried out where the saponification of the ester could be monitored. The signals of the ethyl group of the ester at 4.1 and 1.2 ppm disappear after cleaving the ester in accordance to the literature.[1] Similar to the ammonium bromide polymer p((DMA-Br)-NDI-gT2), the zwitterion polymer aggregated in solution and showed an increased, most likely overestimated, molecular weight distribution (p(ZI-NDI-gT2) $M_n$ = 24 kDa, $M_w$ = 53 kDa). Unfortunately, MALDI-ToF measurements were inconclusive, it was not possible to ionize or detect the ammonium bromide or zwitterion NDI copolymer. It is assumed that the molecular weight of both measured in DMF is overestimated, most likely by a factor of 10.

# 4.Monomer synthesis

## Synthesis of (DMA)-NDI-Br$_2$

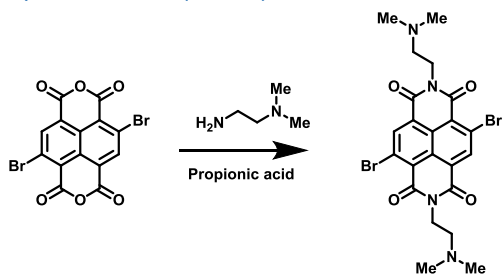

A 150 ml two neck round bottom flask was dried and purged with argon. 2,6-dibromonaphthalene-1,4,5,8-tetracarboxylic dianhydride (512 mg , 1.20 mmol, 1.0 eq.) was suspended in propionic acid (20 ml) and *N,N*-dimethylethane-1,2-diamine (0.23 mg, 2.33 mmol) was added. The reaction mixture was heated to 120°C for 2 h. The reaction was monitored by NMR and after full conversion of the starting material, 100 mL of water and 100 mL of chloroform were added (product is water soluble). The aqueous layer was first washed with chloroform (2 x 100 mL). Then, 200 mL of chloroform was added and a 2 M NaHCO$_3$ solution was added until a pH value of 8 - 9 was reached. The organic layer was washed with water (3 x 100 mL) and dried over MgSO4. The solvent was removed and the red solid was washed with methanol and acetone. Finally, the solid was washed with hot acetone and dried to obtain 250 mg (0.44 mmol) of an orange solid with a yield of 37 %.

$^1$H-NMR (400 MHz, trifluoroacetic acid -$d_1$) $\delta$: 9.00 (s, 2H), 4.68 (t, 4H), 3.68 (t, 4H), 3.13 (s, 12H) ppm. $^{13}$C-NMR (100 MHz, trifluoroacetic acid-$d_1$) $\delta$: 165.5, 142.5, 132.5, 130.2, 127.0, 126.6, 60.42, 46.3, 38.9 ppm. HRMS (ES-ToF): 565.0090 [M-H$^+$] (calc. 565.0086).

Figure S1: $^1$H NMR spectrum of (DMA)-NDI-Br$_2$ measured in TFA-$d_1$.

Figure S2: $^{13}$C NMR spectrum of (DMA)-NDI-Br$_2$ measured in TFA-$d_1$.

## Synthesis of 3,3'-bisalkoxy(TEG)-2,2'-bithiophene

A 250 mL two neck RBF was dried and purged with argon. 3,3'-Dibromo-2,2'-bithiophene (6.48 g, 20 mmol), triethylene glycol monomethyl ether (11.5 g, 70 mmol), $^ت$BuOK (6.73 g, 60 mmol) and CuI (1.52 g, 8.0 mmol) were suspended in anhydrous tert-butanol (100 mL). The reaction mixture was degassed with argon for 15 min and heated to reflux with stirring for 16 h. After the reaction was finished, water was added and the aqueous layer was extracted with ethyl acetate (200 mL). The organic phase was washed with DI water (3 x 100 mL), dried over MgSO$_4$ and concentrated *in vacuo* to afford the crude product as a dark yellow oil. Purification by column chromatography on silica gel with a solvent mixture of ethyl acetate : hexane 3:1 with 1 % of triethylamine afforded the target molecule as a yellow waxy solid (3.2 g, 6.52 mmol, 33% yield).

$^1$H NMR (400 MHz, CDCl$_3$) δ: 7.08 (d, *J* = 5.6 Hz, 2H), 6.85 (d, *J* = 5.6 Hz, 2H), 4.25 (t, *J* = 4.9 Hz, 4H), 3.90 (t, *J* = 5.0 Hz, 4H), 3.78 – 3.75 (m, 4H), 3.69 – 3.66 (m, 4H), 3.66 – 3.64 (m, 4H), 3.55 – 3.53 (m, 4H), 3.37 (s, 6H) ppm. $^{13}$C NMR (100 MHz, CDCl$_3$) δ: 151.9, 122.0, 116.7, 114.9, 72.1, 71.5, 71.1, 70.9, 70.7, 70.2, 59.2 ppm. HRMS (ES-ToF): 491.1774 [M-H$^+$] (calc. C$_{22}$H$_{35}$O$_8$S$_2$ 491.1773).

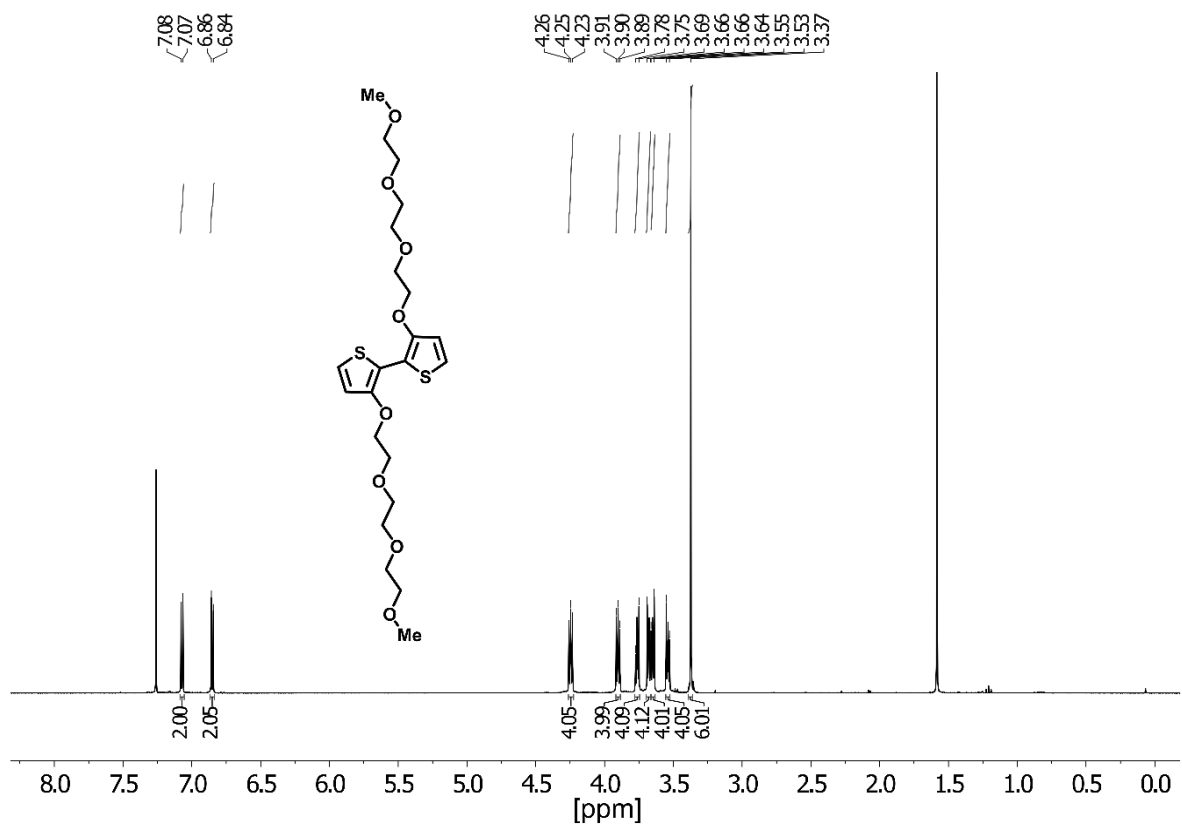

Figure S3: $^1$H NMR spectrum of 3,3'-bisalkoxy(TEG)-2,2'-bithiophene measured in CDCl$_3$.

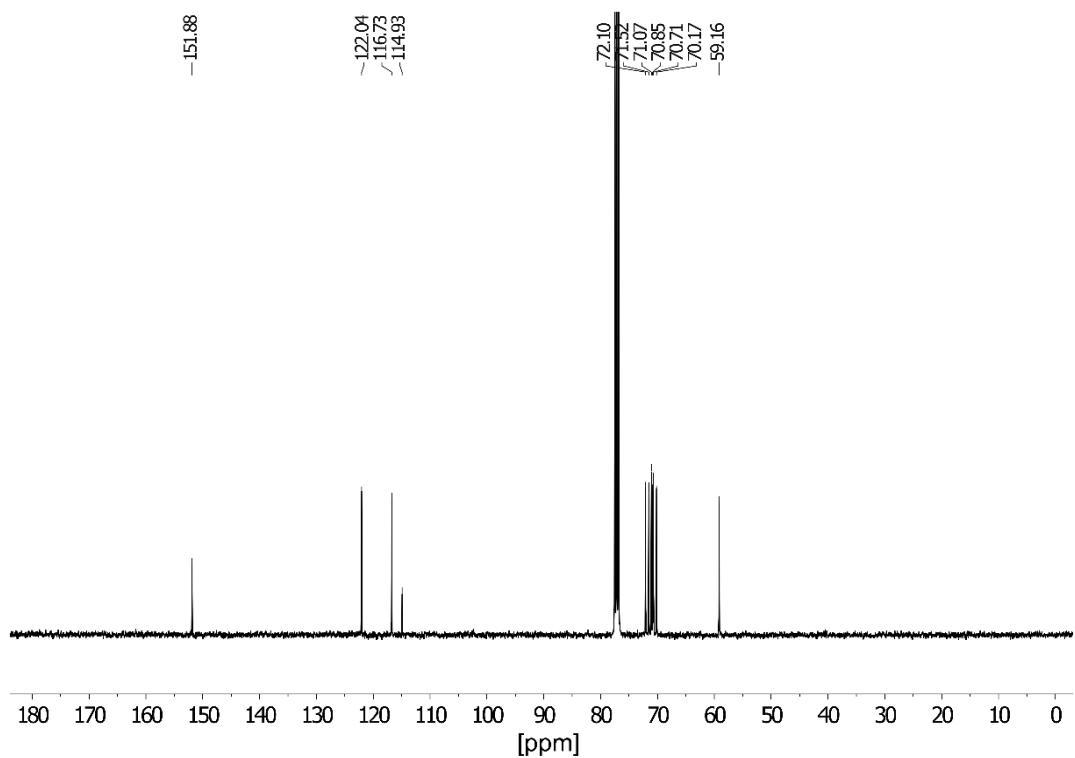

Figure S4: $^{13}$C NMR spectrum of 3,3'-bisalkoxy(TEG)-2,2'-bithiophene measured in CDCl$_3$.

### Synthesis of 5,5'-dibromo-3,3'- bisalkoxy(TEG)-2,2'-bithiophene

A protocol by Nielsen *et al.* was followed and slight modifications were carried out.[2] A two neck RBF was dried under vacuum and purged with argon. 3,3'-bisalkoxy(TEG)-2,2'-bithiophene (1.83 g. 2.24 mmol) was dissolved in anhydrous THF (200 mL), degassed with argon and cooled to - 20 °C. NBS (0.83 g, 4.66 mmol) was added in the dark and the reaction mixture was stirred for 10 min (reaction control indicated full conversion after 10 min). The reaction was quenched by the addition of 100 mL of 1 M NaHCO$_3$ aqueous solution, followed by the addition of 150 mL of ethyl acetate. The organic layer was washed with water (3 x 100 mL), dried over MgSO$_4$ and the solvent was removed under reduced pressure. Purification of the crude product was carried out by column chromatography on silica gel with a solvent mixture of hexane : ethyl acetate in the ration of 1 : 1 with 1 % of triethylamine. Finally, the product was recrystallised from diethyl ether/hexane to afford the product as a yellow solid with a yield of 60 %. (1.45 g, 2.24mmol).

$^1$H NMR (400 MHz, CDCl$_3$) δ: 6.85 (s, 2H), 4.21 – 4.18 (m, 4H), 3.88 – 3.85 (m, 4H), 3.75 – 3.73 (m, 4H), 3.70 – 3.68 (m, 4H), 3.68 – 3.65 (m, 4H), 3.57 – 3.54 (m, 4H), 3.38 (s, 6H) ppm. $^{13}$C NMR (100 MHz, CDCl$_3$) δ: 150.3, 119.8, 116.0, 111.2, 72.1, 71.8, 71.1, 70.9, 70.8, 70.3, 59.2 ppm. HRMS (ES-ToF): 646.9964 [M-H$^+$] (calc. C$_{22}$H$_{33}$Br$_2$O$_8$S$_2$ 646.9983).

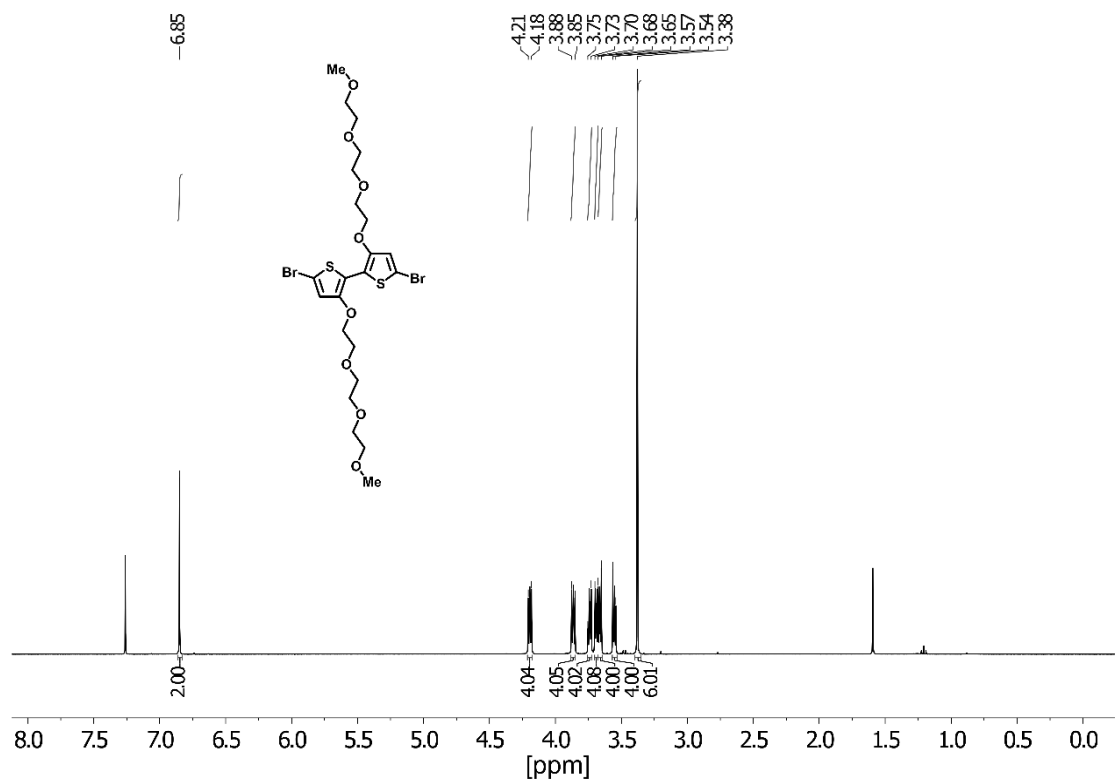

Figure S5: $^1$H NMR spectrum of 5,5'-dibromo-3,3'- bisalkoxy(TEG)-2,2'-bithiophene measured in CDCl$_3$.

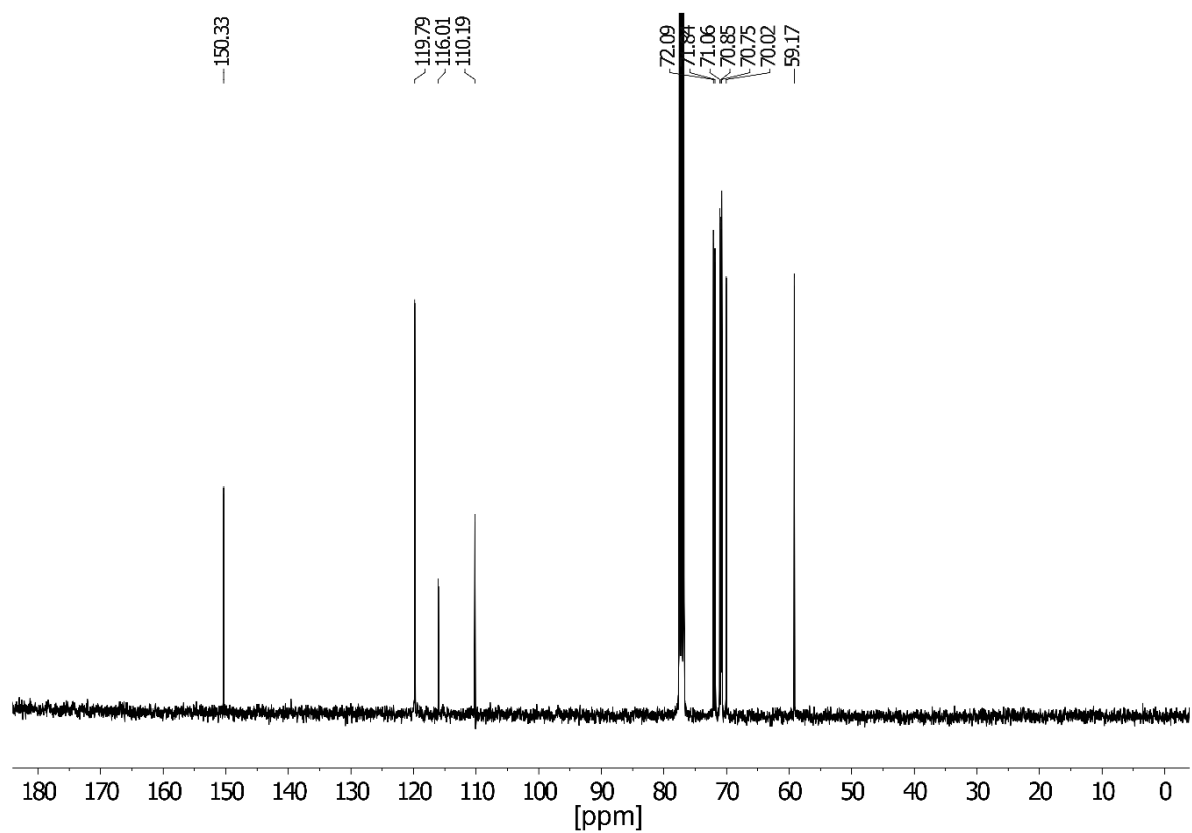

Figure S6: $^{13}$C NMR spectrum of 5,5'-dibromo-3,3'- bisalkoxy(TEG)-2,2'-bithiophene measured in CDCl$_3$.

## Synthesis of (3,3'- bisalkoxy(TEG)-[2,2'-bithiophene]-5,5'-diyl)bis(trimethylstannane)

A 250 mL two neck RBF was dried and purged with argon. 1.01 g of 5,5'-dibromo-3,3'- bisalkoxy(TEG)-2,2'-bithiophene (1.56 mmol, 1.0 eq.) was dissolved in 150 mL of anhydrous THF. The reaction mixture was cooled to -78 °C and 2.5 mL of *n*-BuLi (2.5 M in hexane, 6.23 mmol, 4.0 eq.) was added slowly. A yellow solution was formed which was stirred at -78 °C for 3 h. Then, 7.8 mL of trimethyltin chloride (1 M in hexane, 7.8 mmol, 5.0 eq.) was added and the reaction mixture was warmed to room temperature. 150 mL of diethyl ether was added and the organic phase was washed with water (3 x 100 mL) and dried over Na$_2$SO$_4$. The solvent was removed and the obtained solid was dissolved in 100 mL of acetonitrile which was washed with hexane (3 x 100 mL). The solvent was removed and the product was recrystallized from diethyl ether to obtain the product as yellow needles with a yield of 64 % (810 mg, 0.99 mmol).

$^1$H NMR (400 MHz, acetone-*d$_6$*) δ: 8.99 (s, 2H), 4.28 – 4.26 (m, 4H), 3.88 – 3.85 (m, 4H), 3.71 – 3.69 (m, 4H), 3.62 – 3.56 (m, 8H), 3.46 – 3.44 (m, 4H), 3.27 (s, 6H), 0.37 (s, 18 H) ppm. $^{13}$C NMR (100 MHz, acetone-*d$_6$*): 154.9, 134.9, 125.3, 121.1, 72.8, 72.4, 71.8, 71.3, 71.0, 59.0, 8.24 ppm. HRMS (ES-ToF): 819.1069 [M-H+] (calc. 819.1084).

Figure S7: ¹H NMR spectrum of (3,3'- bisalkoxy(TEG)-[2,2'-bithiophene]-5,5'-diyl)bis(trimethylstannane) in acetone-$d_6$.

Figure S8: ¹³C NMR spectra of (3,3'- bisalkoxy(TEG)-[2,2'-bithiophene]-5,5'-diyl)bis(trimethylstannane) in acetone-$d_6$.

# 5. Polymer synthesis

## Synthesis of poly(TEG-alkoxybithiophene) p(gT2)

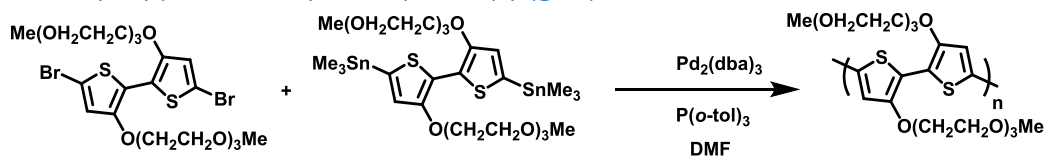

In a dried 5.0 mL microwave vial, 79.4 mg of (3,3'- bisalkoxy(TEG)-[2,2'-bithiophene]-5,5'-diyl)bis(trimethylstannane) (97.3 µmol) and 63.1 mg of 5,5'-dibromo-3,3'- bisalkoxy(TEG)-2,2'-bithiophene (97.3 µmol) were dissolved in 3.0 mL of anhydrous, degassed DMF. $Pd_2(dba)_3$ (1.78 mg, 1.95 µmol) and P(*o*-tol)$_3$ (2.37 mg, 7.78 µmol) were added and the vial was sealed and heated to 100 °C for 16 h. After the polymerization has finished, the end-capping procedure was carried out. Then, the reaction mixture was cooled to room temperature and precipitated in methanol. A blue solid was formed which was filtered into a glass fibre-thimble and Soxhlet extraction was carried out with methanol, ethyl acetate, acetone, hexane, and chloroform. The polymer dissolved in hot chloroform. Finally, the polymer was dissolved in a minimum amount of chloroform and precipitated in methanol. The collect solid was filtered and dried under high vacuum. A blue solid was obtained with a yield of 74 % (70 mg, 71.6 µmol).

GPC (DMF, 50 °C) $M_n$ = 56 kDa, $M_w$ = 306 kDa. $^1$H NMR (CDCl$_3$, 400 MHz) δ: 6.96 (br s, 2 H), 4.35 (br s, 4 H), 4.01 – 3.92 (m, 4 H), 3.81 – 3.79 (m, 4 H), 3.72 – 3.70 (m, 4 H), 3.66 – 3.63 (m, 4 H), 3.53 – 3.50 (m, 4 H), 3.34 (s, 6 H) ppm.

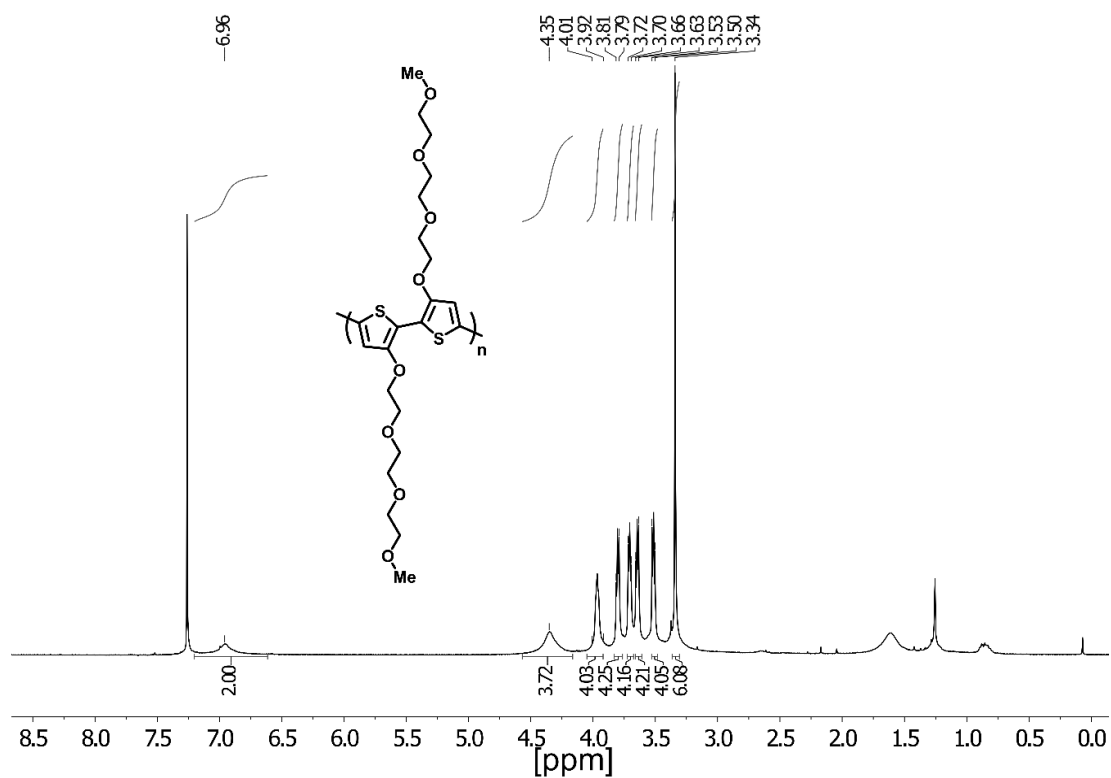

Figure S9: $^1$H NMR spectrum of p(gT2) in CDCl$_3$ at 25 °C.

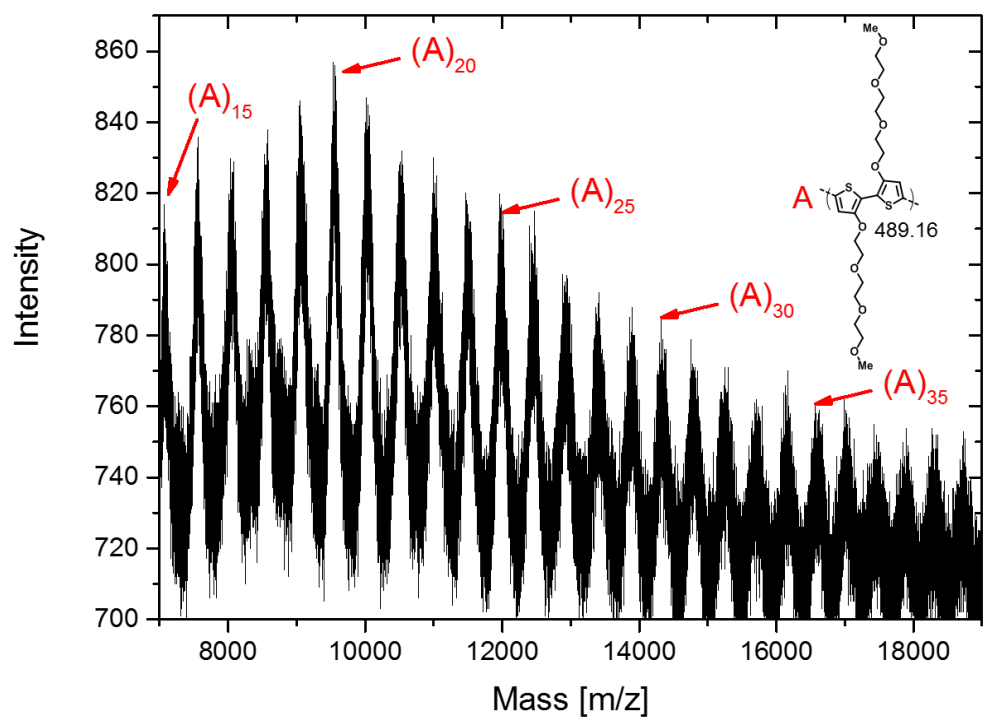

Figure S10: MALDI-ToF spectrum of p(gT2), measured in positive linear mode with DCTB as the matrix.

## Synthesis of p((DMA)-NDI-gT2)

In a 5.0 mL microwave vial, 38.76 mg of (DMA)-NDI-Br$_2$ (68.45 µmol, 1.0 eq.), 55.86 mg of (3,3'-bisalkoxy(TEG)-[2,2'-bithiophene]-5,5'-diyl)bis(trimethylstannane) (68.45 µmol, 1.0 eq.), 1.4 mg of Pd$_2$(dba)$_3$ (1.52 µmol, 2 mol%) and 1.87 mg of P(o-tol)$_3$ (6.12 µmol, 8 mol%) were dissolved in 1.5 mL of anhydrous, degassed DMF and the vial was heated to 85 °C for 16 h. The color changed from yellow to dark green. After the polymerization has finished, the end-capping procedure was carried out. Then, the reaction mixture was cooled to room temperature and the reaction mixture was precipitated in ethyl acetate. The precipitate was filtered and Soxhlet extraction was carried out with ethyl acetate, methanol, acetone, hexane and chloroform. The polymer was soluble in hot chloroform. Polymer p((DMA)-NDI-gT2) was obtained as a green solid with a yield of 79 % (48.4 mg, 54.1 µmol).

GPC (CHCl$_3$, 50 °C) M$_n$ = 5.3 kDa, M$_w$ = 8.9 kDa. $^1$H-NMR (400 MHz, CHCl$_3$) δ: 8.82 (br s, 2H), 7.26 (br s, 4H), 4.42 (br s, 4H), 4.35 (br s, 4H), 4.26 (br s, 4H), 3.91 – 383 (m, 4H), 3.83 – 3.75 (m, 4H), 3.75 – 3.64 (m, 4H), 3.64 – 3.54 (m, 4H), 3.54 – 3.45 (m, 4H), 3.39 – 3.28 (m, 6H), 2.69 – 2.58 (m, 4H), 2.40 – 2.22 (m, 12H) ppm.

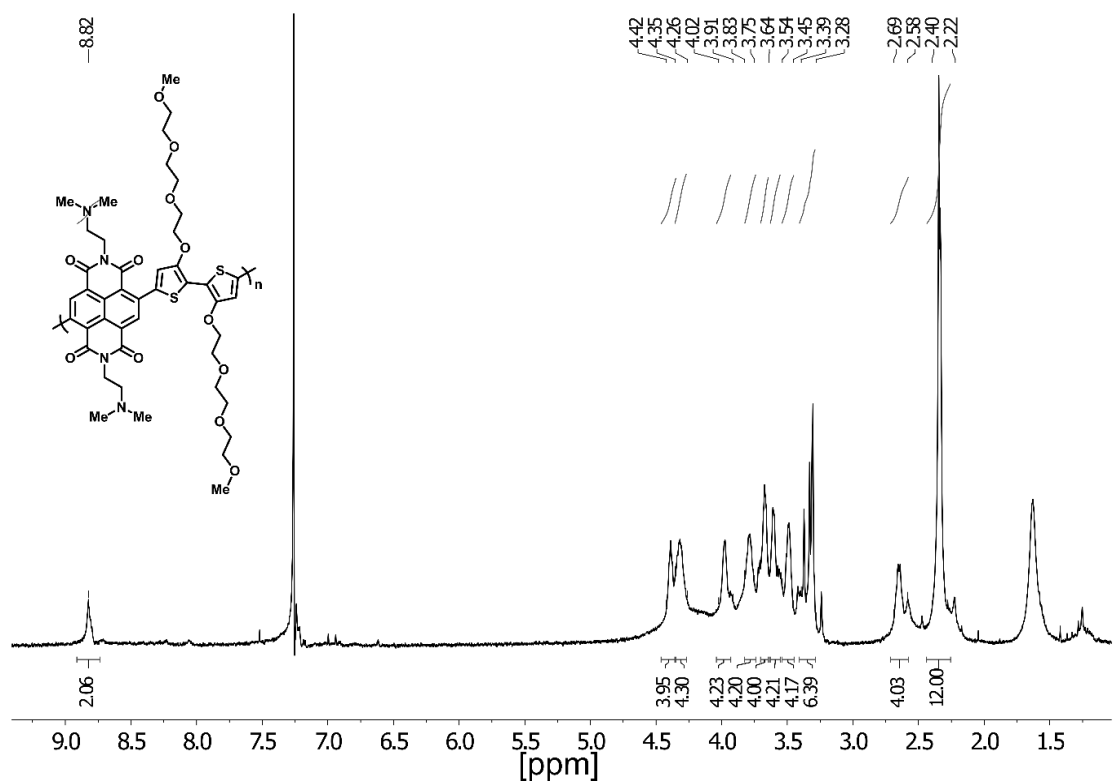

Figure S11: $^1$H NMR spectrum of p((DMA)-NDI-gT2) in CDCl$_3$ at 25 °C.

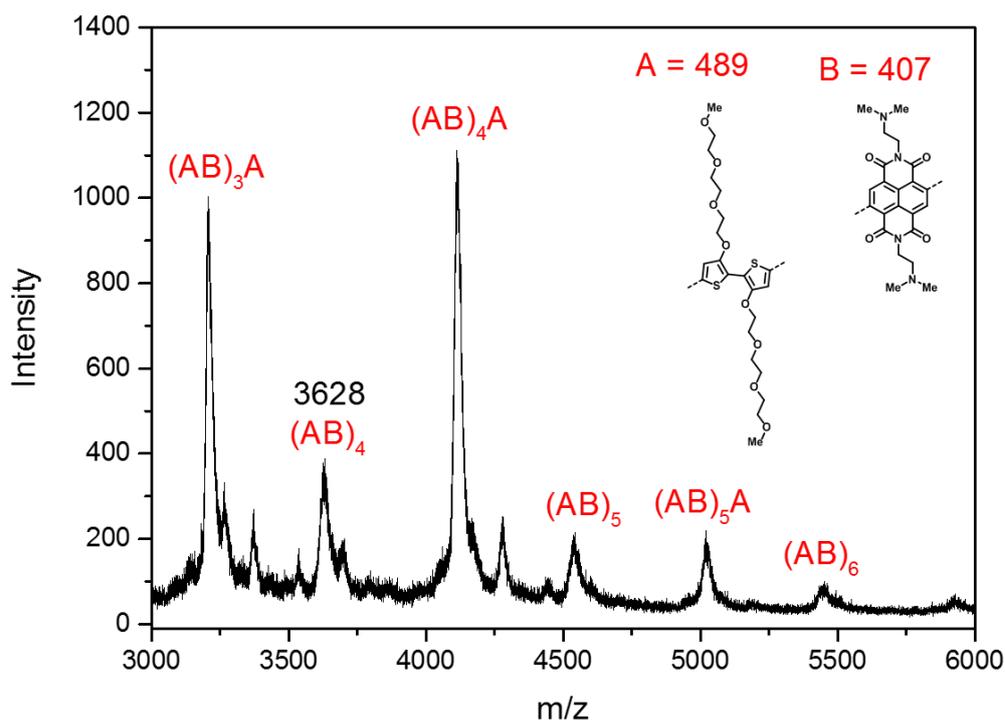

Figure S12: MALDI-ToF spectrum of p((DMA)-NDI-gT2).

## Synthesis of p((DMA-Br)-NDI-gT2)

A 5.0 mL microwave vial was dried and purged with argon, 14.3 mg p((DMA)-NDI-gT2) (15.9 µmol) was suspended in 4.00 ml anhydrous DMF and 0.5 ml ethyl-6-bromohexanoate (2.8 mmol) was added. The reaction mixture was heated to 120°C for 1.5 h and a green solution was formed. The reaction mixture was cooled to room temperature and the solvent was removed under reduced pressure. The green solid was suspended in 3 mL of methanol and precipitated in acetone followed by the addition of hexane. The solution was filtered and the green solid was washed with chloroform and acetone. Finally, the polymer was dried under high vacuum for 16 h. 20.6 mg (15.4 µmol) of the polymer was obtained as a green solid with a yield 97%.

GPC (DMF, 50 °C) $M_n$ = 38 kDa, $M_w$ = 60 kDa. $^1$H-NMR (500 MHz, DMSO-$d_6$) δ: 8.64 – 8.57 (m, 2H), 7.81 – 7.48 (m, 4H), 4.49 – 4.24 (m, 8H, signals overlapping), 4.07 – 3.98 (m, 4H, COCH$_2$), 3.98 – 3.89 (m, 4H), 3.68 – 3.62 (m, 4H), 3.57 – 3.50 (m, 8H), 3.47 – 3.40 (m, 8H), 3.22 – 3.11 (m, 12H), 2.95 – 2.86 (m, 4H), 3.98 – 3.89 (m, 4H), 2.37 – 2.29 (m, 4H), 1.80 – 1.70 (m, 4H), 1.62 – 1.54 (m, 4H), 1.37 – 1.27 (m, 4H), 1.18 – 1.13 (m, 6H) ppm.

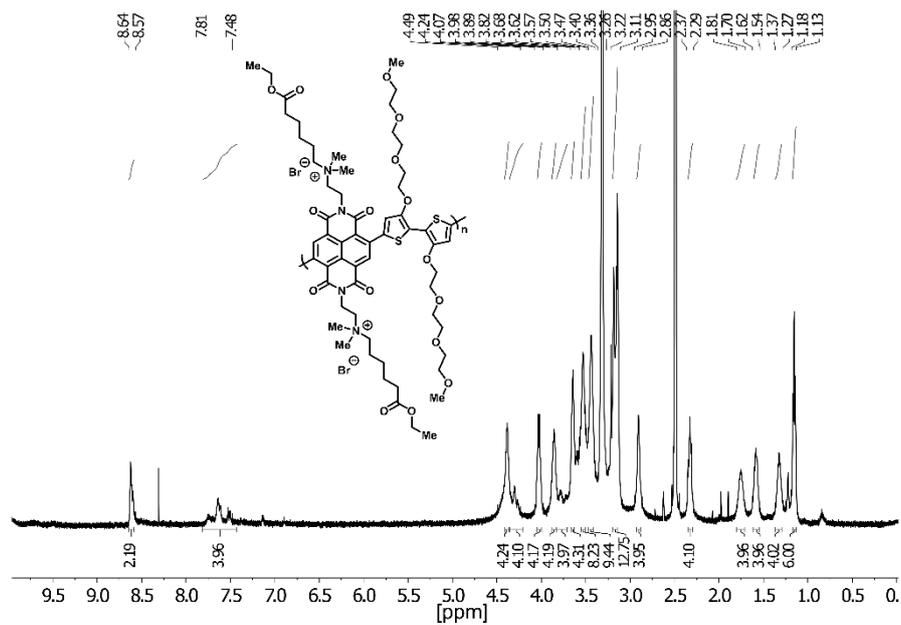

Figure S13: ¹H NMR spectrum of p((DMA-Br)-NDI-gT2) in DMSO-$d_6$ at 22 °C.

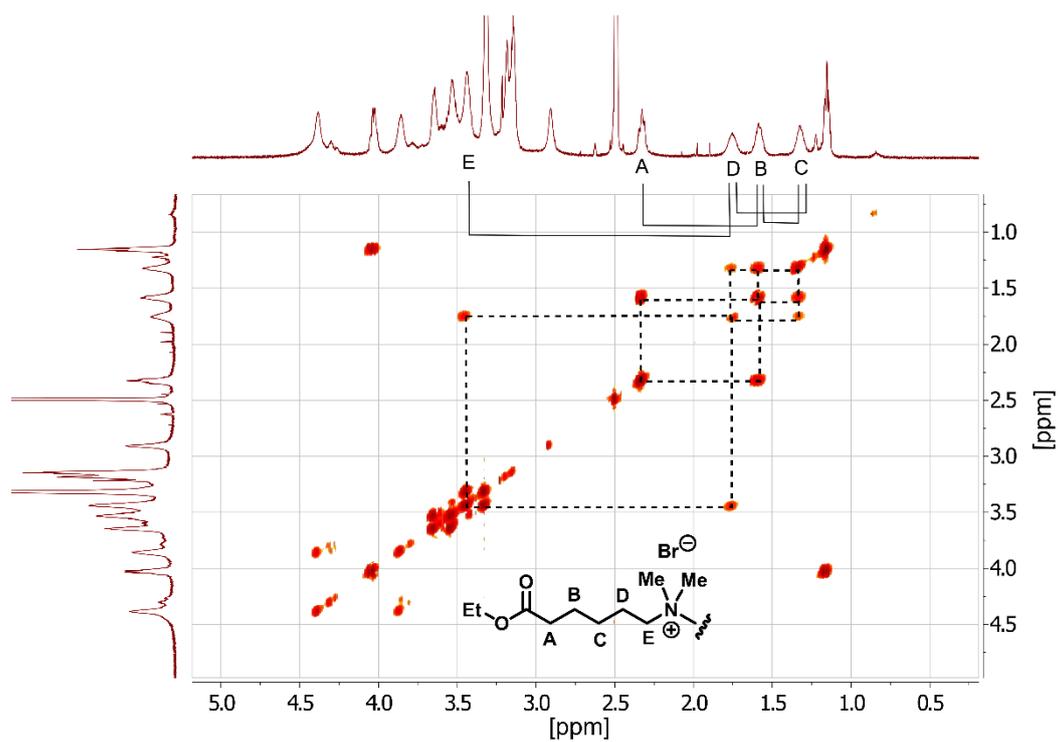

Figure S14: 2D COSY spectrum of p((DMA-Br)-NDI-gT2) in DMSO-$d_6$ at 25 °C, proton couplings of the alkylation are highlighted.

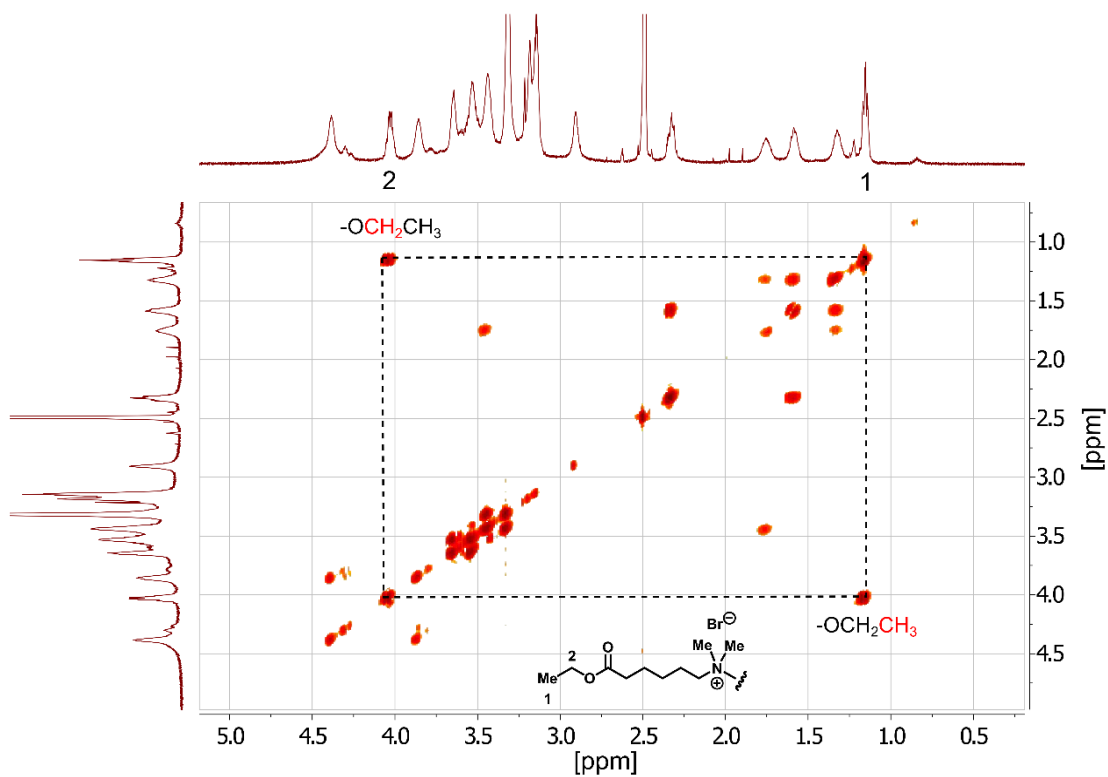

Figure S15: 2D COSY spectrum of p((DMA-Br)-NDI-gT2) in DMSO-$d_6$ at 25 °C, proton couplings of ester group are highlighted.

## Synthesis of p(ZI-NDI-gT2)

In a 7.0 mL microwave, 10 mg of p((DMA-Br)-NDI-gT2) (8.46 µmol, 1.00 eq.) was dissolved in 2 mL of DMSO and 2 mL of water. 0.1 mL of conc. HCl was added and the reaction mixture was heated to 75 °C for 16 h. Then, the reaction mixture was cooled to room temperature and the solution was transferred into a dialyses kit (molecular weight cut off 2kDa) and the dialysis kit was stirred in DI water for 2 days, exchanging the water every 6 h. Finally, solvent was removed and the polymer was dried at 60 °C for 16 h. 9.2 mg (8.19 µmol) of a green polymer was obtained with a yield of 97 %.

GPC (DMF, 50 °C) $M_n$ = 24 kDa, $M_w$ = 53 kDa. $^1$H-NMR (500 MHz, DMSO-$d_6$) δ: 8.67 – 8.55 (m, 2H), 7.73 – 7.45 (m, 4H), 4.48 – 4.30 (m, 8H, signals overlapping), 3.61 – 3.56 (m, 8H, signals overlapping), 3.55 – 3.52 (m, 4H), 3.46 – 3.40 (m, 8H), 3.40 – 3.24 (m, 4H, overlap with water peak), 3.23 – 3.12 (m, 12H), 2.30 – 2.21 (m, 4H), 1.81 – 1.70 (m, 4H), 1.62 – 1.54 (m, 4H), 1.37 – 1.28 (m, 4H) ppm.

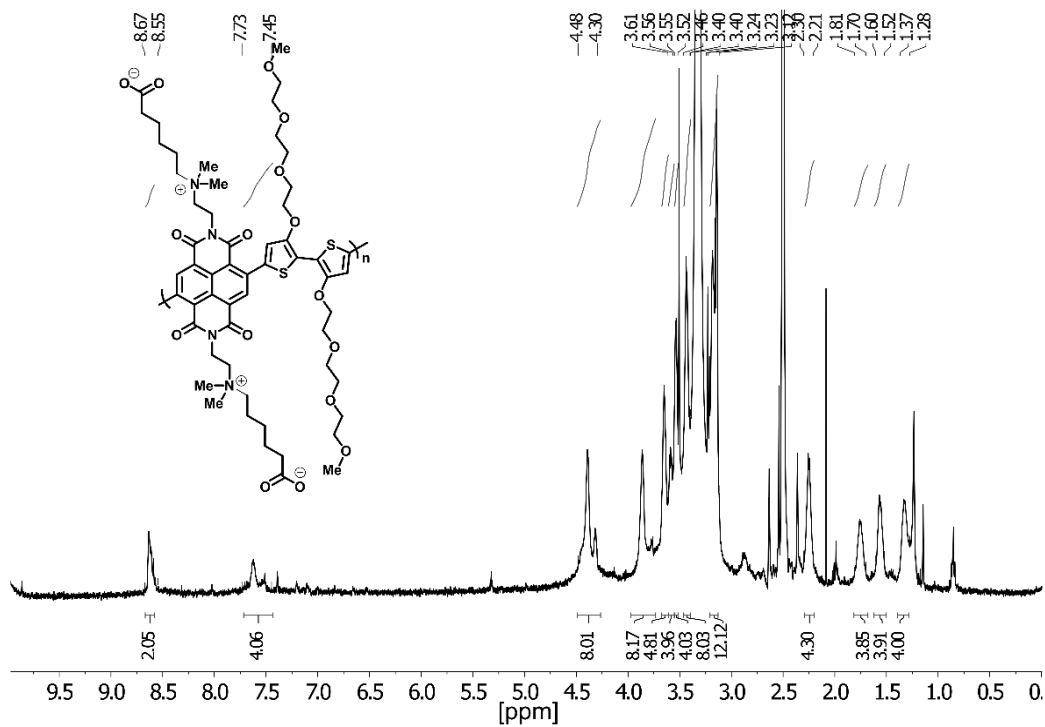

Figure S16: ¹H NMR spectrum of p(ZI-NDI-gT2) in DMSO-*d₆*.

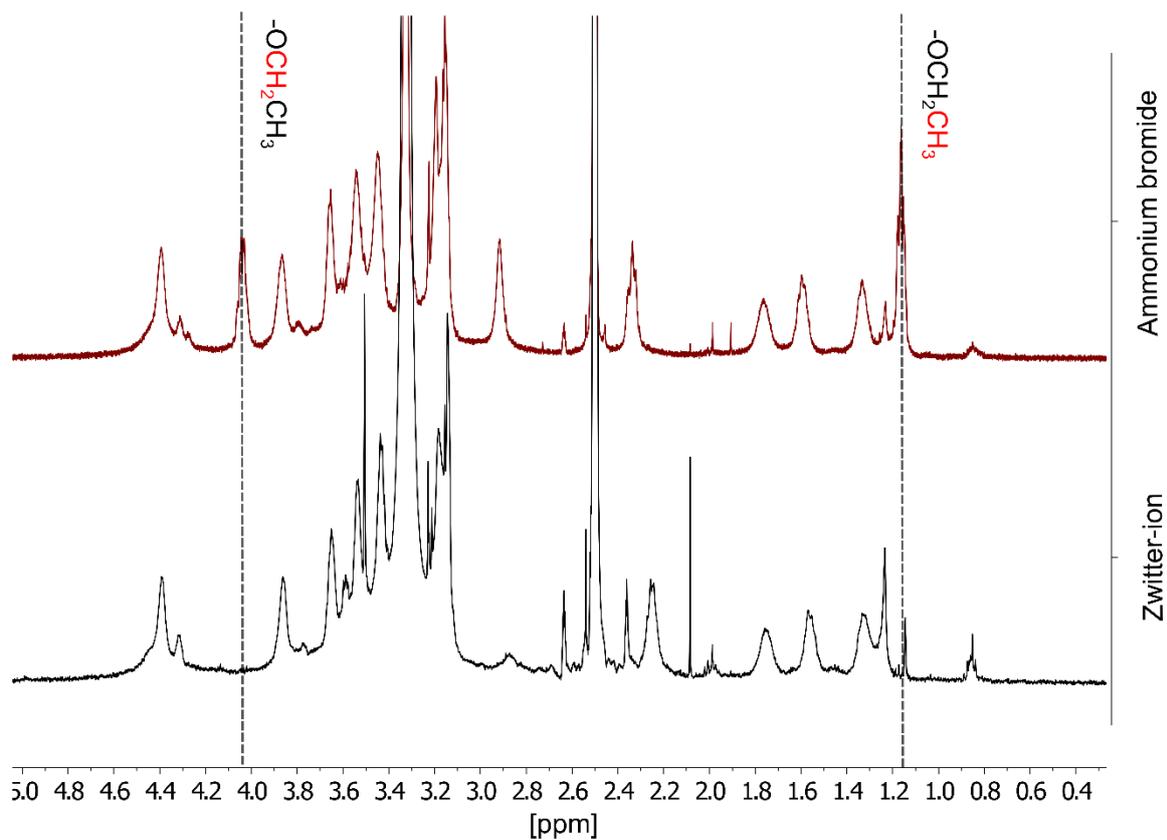

Figure S17: ¹H NMR spectrum of the ammonium bromide ester p((DMA-Br)-NDI-gT2) and zwitterion p(ZI-NDI-gT2) in DMSO-*d₆*. The disappearance of the ester group is highlighted.

## Synthesis of p(g7NDI-gT2)

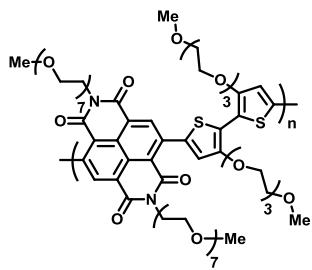

In a 7.0 mL microwave vial, 72.25 mg of g7NDI-Br$_2$ (67.6 µmol, 1.0 eq.) and 55.17 mg of (3,3'-bisalkoxy(TEG)-[2,2'-bithiophene]-5,5'-diyl)bis(trimethylstannane) (67.6 µmol, 1.0 eq.) were dissolved in 2.0 mL of anhydrous, degassed chlorobenzene. 1.36 mg of Pd$_2$(dba)$_3$ (1.35 µmol, 2 mol%) and 1.78 mg of P(*o*-tol)$_3$ (5.4 µmol, 8 mol%) were added and the vial was heated to 135 °C for 16 h. After the polymerization has finished, the end-capping procedure was carried out. Then, the reaction mixture was cooled to room temperature and the dark green reaction mixture was precipitated in ethyl acetate followed by addition of hexane. Soxhlet extraction was carried out with hexane, ethyl acetate, MeOH, acetone, THF and chloroform. The polymer was soluble in hot chloroform. Polymer p(g7NDI-gT2) was obtained as a dark green solid with a yield of 76 % (48 mg, 0.05 mmol).

GPC (CHCl$_3$:DMF; 5:1, 50 °C) M$_n$ = 14 kDa, M$_w$ = 26 kDa. $^1$H NMR (400 MHz, CDCl$_3$) δ: 8.82 (s, 2H), 7.31 – 7.16 (m, 4H), 4.44 – 4.34 (m, 4H), 4.34 – 4.25 (m, 4H), 4.02 – 3.93 (m, 4H), 3.89 – 3.75 (m, 4H), 7.31 – 7.16 (m, 4H), 3.70 – 3.52 (m, 60H), 3.52 – 3.49 (m, 4H), 3.36 (br s, 12H).

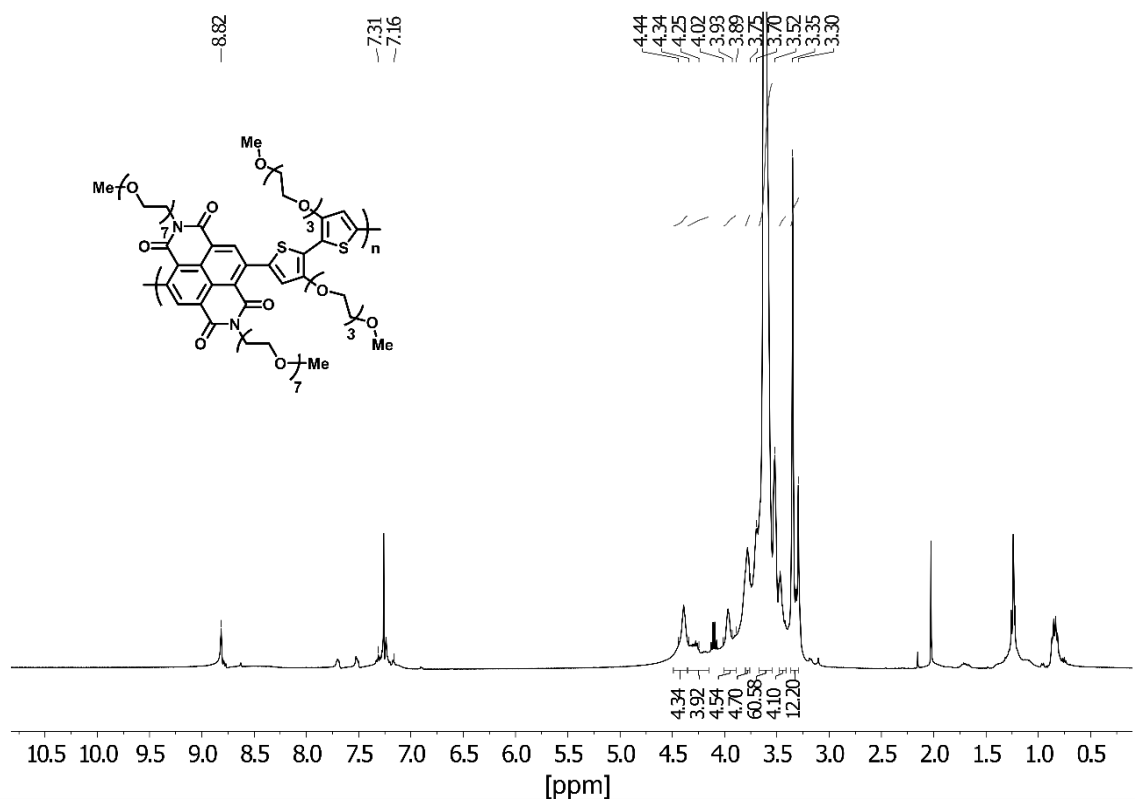

Figure S18: $^1$H NMR spectrum of polymer p(g7NDI-gT2) in CDCl$_3$ at 22 °C, aromatic signals were not integrated due to the overlap with the solvent peak.

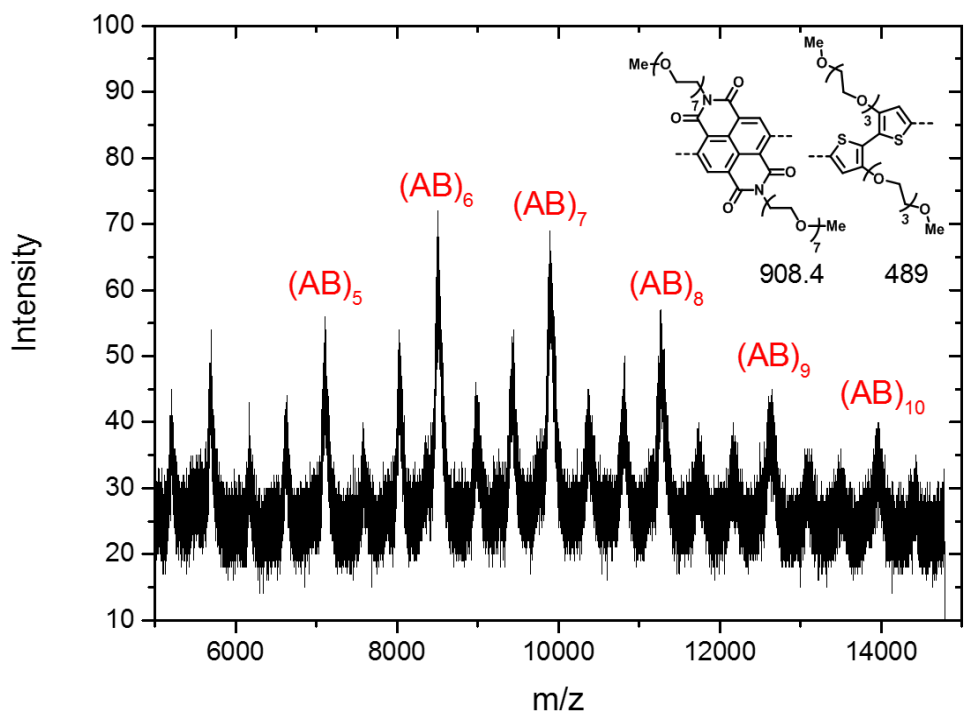

Figure S19: MALDI-ToF mass spectrum of p(g7NDI-gT2).

# 6. CV measurements in acetonitrile

Thin film CV measurements were carried out on ITO or FTO coated glass substrates with a 0.1 M TBAPF$_6$ acetonitrile solution as the supportive electrolyte. Potentials are reported vs Ag/AgCl and the scan rate of the measurement was 100 mV/s. The oxidation of ferrocene (solution) was used as the reference to calculate IP and EA values.

## p(gT2)

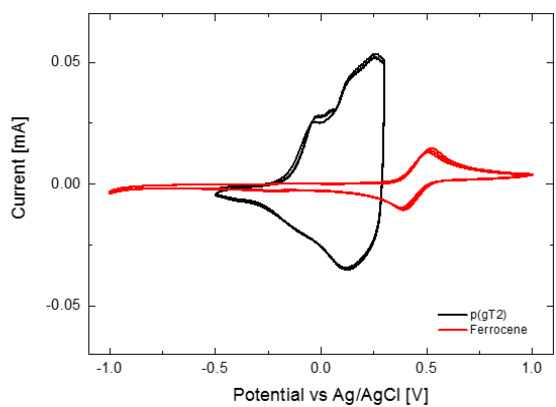

Figure S20: CV measurement of p(gT2) and ferrocene.

## p(g7NDI-gT2)

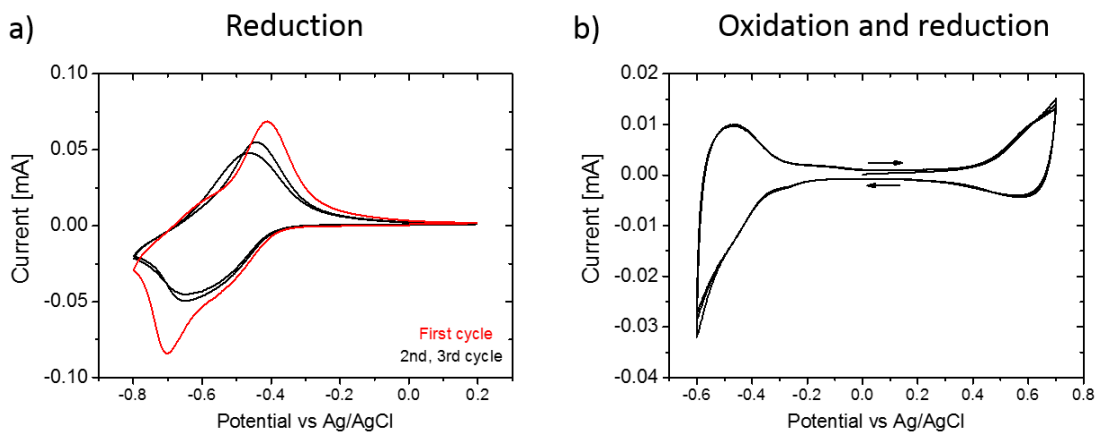

Figure S21: CV measurement of p(g7NDI-gT2) (a) reduction only and (b) oxidation and reduction (reversible regime).

p((DMA)-NDI-gT2), p((DMA-Br)-NDI-gT2) and p(ZI-NDI-gT2)

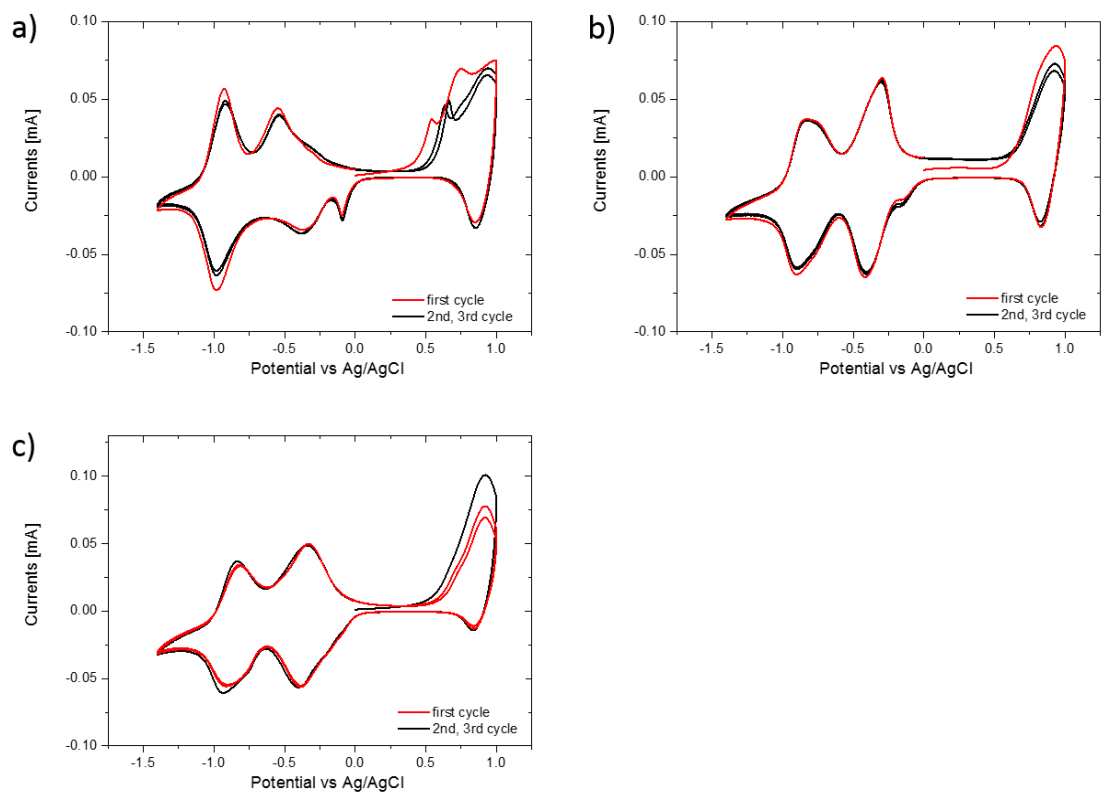

Figure S22: CV measurement of (a) p((DMA)-NDI-gT2), (b) p((DMA-Br)-NDI-gT2) and (c) p(ZI-NDI-gT2).

# 7. UV Vis measurements

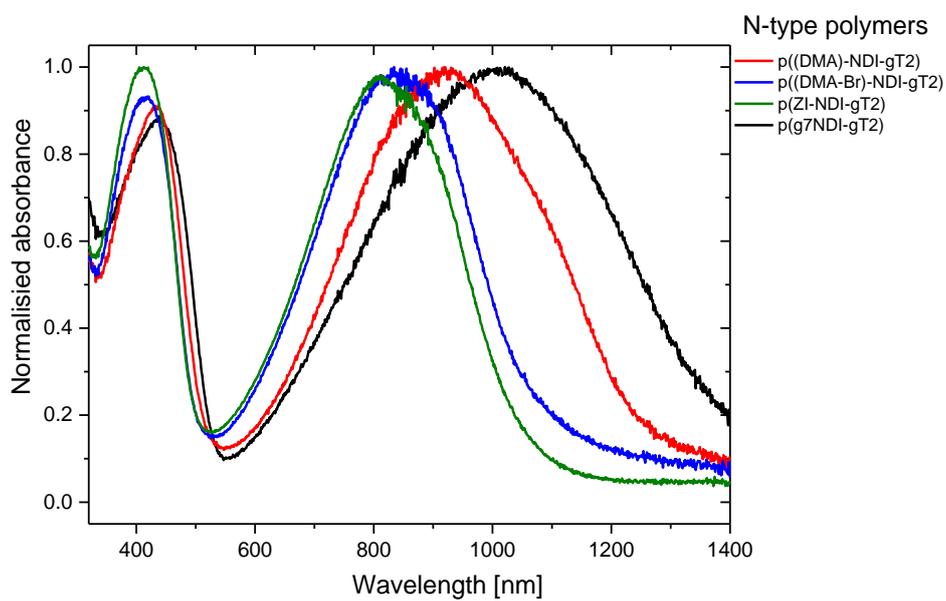

Figure S23: Normalized UV-Vis spectra of the n-type copolymers prepared by spin coating or doctor blade coating on glass substrates.

# 8. Spectroelectrochemical measurements and sample preparation

### Sample preparation

The preparation of samples presented in the main text was carried out according to the following procedure: fluorine doped tin oxide (FTO TEC15) coated glass substrates were cleaned with soap, deionized water acetone and isopropanol before undergoing a heating step at 450°C for 30 minutes. The polymers dissolved in a solvent were then deposited from solution onto the substrates via doctor blading. p(gT2) and p(g7NDI-gT2) were deposited from a chloroform solution at room temperature, while p(ZI-NDI-gT2) was deposited from dimethyl sulfoxide (DMSO) at 160°C. For the p(ZI-NDI-gT2) polymer an additional heating step at 150°C for 10 minutes was performed after the deposition in order to remove residuals of solvent in the film. Araldite rapid epoxy glue was added whenever it was needed to minimize any exposed bare FTO surface to the electrolyte and the sample was left in ambient condition overnight before the electrochemical measurements. The thickness of the films was calculated from optical absorbance of the samples. The measurement of film thickness for a certain material was calibrated against a reference value of absorbance of a sample of that material with known thickness (measured using a Dektak profilometer).

### Electrochemical and spectroelectrochemical measurements

Electrochemical characterization presented in the main text was carried out using an Ivium CompactStat potentiostat in an electrochemical cell filled with 0.1 M NaCl:DIW electrolyte, using the polymer film as the working electrode of the cell, a Ag/AgCl 3 M NaCl reference electrode and a platinum counter electrode. The electrolyte was flushed with argon for at least 15 minutes before the beginning of measurements involving reduction of n-type polymers to avoid reaction of electrons with molecular oxygen. The electrochemical cell was a quartz cuvette with transparent windows which enabled simultaneous optical spectra acquisition through a UV-vis spectrometer (OceanOptics USB 2000+) which collected the transmitted light through the sample from a tungsten lamp used as probe light source. Spectroelectrochemical measurements on a battery cell including both p-type and an n-type polymer films were performed in a similar way, using a 2 electrode configuration and by monitoring the optical absorption of both films in series. Galvanostatic measurements were performed by applying a sequence of charging and discharging constant current levels to the sample and monitoring the voltage of the sample (for measurements of single electrodes) or of the cell (for full battery measurements). The charging and the discharging currents were of same magnitude in all cases. Multiple consecutive cycles were performed to evaluate the rate capabilities of the battery for each value of applied current and the $2^{nd}$ cycle was used to calculate the specific capacity illustrated in Figure 4b. Regarding the cyclic voltammetry characterization, the battery was scanned to a negative potential (between -0.8 V and 1 V) to reset the neutrality of the p-type polymer between scans presented in Figure 4a. This procedure was followed to avoid distortion in the CV profiles due to charge retention unbalance between the cathode and anode electrodes. This effect and its consequences are discussed in details in Supplementary Section 19.

# 9. DFT and TD-DFT simulations

## Method

Molecules in their neural, singly and doubly reduced (for n-type) or oxidized (for p-type) states were optimized at the DFT level (B3LYP/6-31g(d,p)). TD-DFT calculations were then performed on the optimized geometries (B3LYP/6-31g(d,p)) to obtain absorption spectra. All quantum chemical calculations were done using Gaussian16.[3]

## Results and discussion

**p-type polymer - p(gT2)**

In order to reproduce experimental data on p(gT2), oligomers (gT2)$_n$ with n=1, 3 and 6 were calculated using above described procedure. For neutral systems, the lowest energy absorption peak position red-shifts with increasing number of monomer units from 328 nm for monomer, 583 nm for trimer to 763 nm for hexamer. These spectra are presented in Figure S24. The experimental value is around 645 nm as shown in Figure 2 of the main text. On this basis, we chose the trimer - (gT2)$_3$ as a reference molecule to compare with experimental results (the chemical structure of (gT2)$_3$ is shown in Figure S25). Figure S26 presents calculated spectra of (gT2)$_3$ in its neutral ('0'), singly ('+1') and doubly ('+2') oxidized states with Gaussian broadening equal 0.15 eV. Bottom panel of Figure S26 presents different linear combinations of these three spectra to illustrate some of the expected intermediate states. Table S1 presents calculated active excited states energies and corresponding oscillator strengths of (gT2)$_3$, (gT2)$_3^{+1}$ and (gT2)$_3^{+2}$.

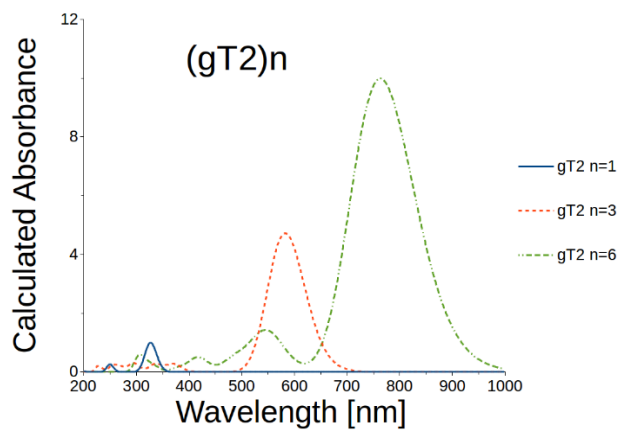

Figure S24. TD-DFT calculated absorption spectra of the monomer, trimer and hexamer (gT2)$_n$ in their neutral state normalized to the spectra of monomer.

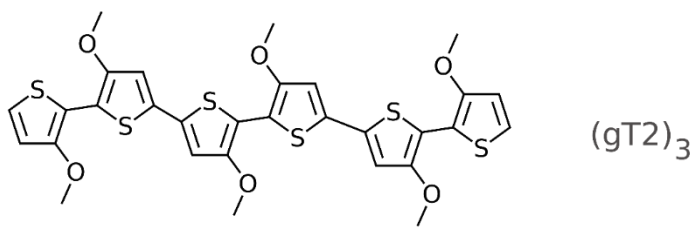

Figure S25. Chemical structure of the trimer (gT2)$_3$.

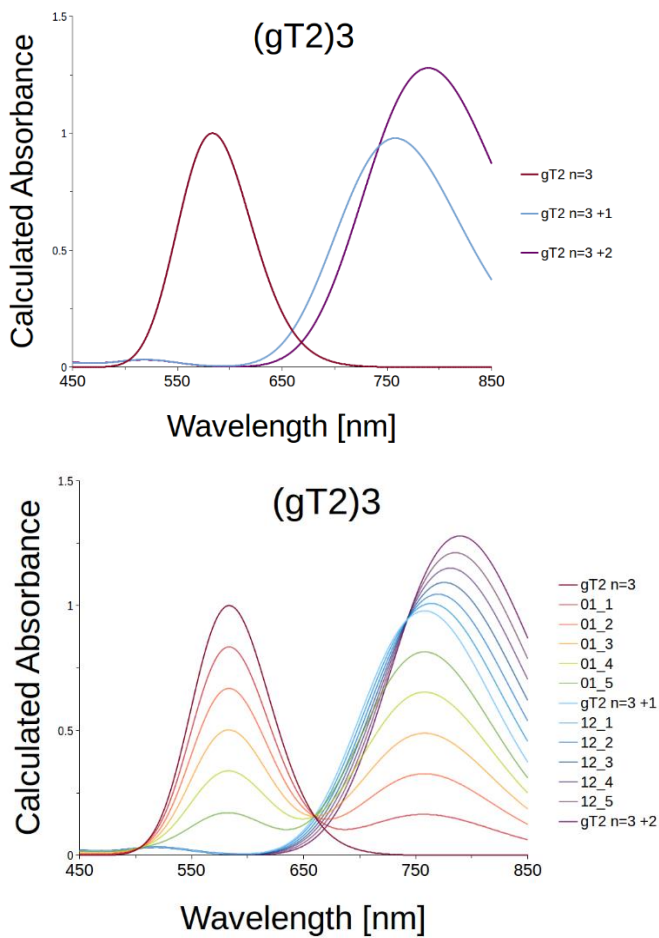

Figure S26. (top) TD-DFT calculated absorption spectra of the trimer (gT2)₃ in its neutral, singly and doubly oxidized state normalized to the spectra of neutral trimer. (bottom) Linear combination of the spectra qualitatively reproducing results shown in the main text for p(gT2). In the legend eg. '01_2' means the second (2) out of five step between neutral (0) and singly oxidized (1) trimer. Other names were created analogously.

Table S1: Energies and oscillator strengths calculated for $(gT2)_3$, $(gT2)_3^{+1}$ and $(gT2)_3^{+2}$.

| Neutral (gT2)₃ | | (gT2)₃⁺¹ | | (gT2)₃⁺² | |
|---|---|---|---|---|---|
| Energy [eV (nm)] | Oscillator strength | Energy [eV (nm)] | Oscillator strength | Energy [eV (nm)] | Oscillator strength |
| 2.12 (583.6) | 2.05 | 0.89 (1383.3) | 0.27 | 1.57 (789.4) | 2.62 |
| 3.19 (388.6) | 0.004 | 1.64 (758.1) | 2.01 | 2.37 (522.5) | 0.06 |
| 3.34 (370.9) | 0.12 | 2.20 (562.7) | 0.003 | 2.57 (481.6) | 0.02 |
| 3.67 (337.2) | 0.12 | 2.37 (522.5) | 0.02 | 2.77 (447.3) | 0.02 |
| | | 2.39 (517.8) | 0.05 | 2.87 (432.2) | 0.01 |
| | | 2.64 (470.4) | 0.01 | 3.01 (412.3) | 0.04 |
| | | 2.76 (449.8) | 0.03 | | |
| | | 3.02 (410.4) | 0.05 | | |

## n-type polymer - p(g7NDI-gT2) and p(ZI-NDI-gT2)

To reduce computational time, n-type polymers p(g7NDI-gT2) and p(ZI-NDI-gT2) were approximated by their monomers with shorter side chains. Figure S27 shows structural formulas of monomers named gT-g1DI-gT and gT-ZI-NDI-gT, both consist of NDI core with ethylene glycol (g1NDI) or zwitterion (ZI-NDI) side chains and methoxy thiophene ring on each side of NDI (gT). Figure S28 and S29 show spectra of monomers in their neutral, singly and doubly reduced states with Gaussian broadening equal 0.25 eV. Bottom panels include spectra of partially charged species. Table S2 and S3 consist of list of excited states and corresponding oscillator strengths for gT-g1NDI-gT and gT-ZI-NDI-gT respectively.

Together with optical spectra, charge distributions across molecules in different oxidation states were calculated using CHEL PG scheme (Charges from Electrostatic Potentials using a Grid-based method) implemented in Gaussian16. In order to simplify the analysis of the results, molecules are divided into fragments, indicated with different colors in Figure S30 and Figure S31. Table S4 and Table S5 show total charge distributions for these fragments in neutral, polaron and bipolaron states of gT-g1NDI-gT and gT-ZI-NDI-gT respectively. Differences between neutral and polaron states and bipolaron and polaron states are also included in the tables.

As we comment in the main text of the paper, most of the added charge is distributed within the NDI core. In the case of gT-g1NDI-gT, 68% of the first extra electron (polaron) and 67% of the second extra electron (bipolaron) are located on the NDI. In the zwitterion case – gT-ZI-NDI-gT, the addition of the first extra electron causes an even stronger polarization in the side chains than in the neutral molecule and thus stronger charge localization on NDI – 85%. Including a second electron already breaks this trend. Only 58% of its charge is localized on NDI, the rest is spread among side chains and thiophenes.

In both molecules, the oxygen atoms in the NDI group are negatively charged in the neutral state and become even more negative upon reduction, much more than their neighboring positively charged carbons. This fact suggests that oxygens are favored locations for extra electrons in the first and second reductions of gT-g1NDI-gT and gT-ZI-NDI-gT and justifies the reduction mechanism proposed in reference [4].

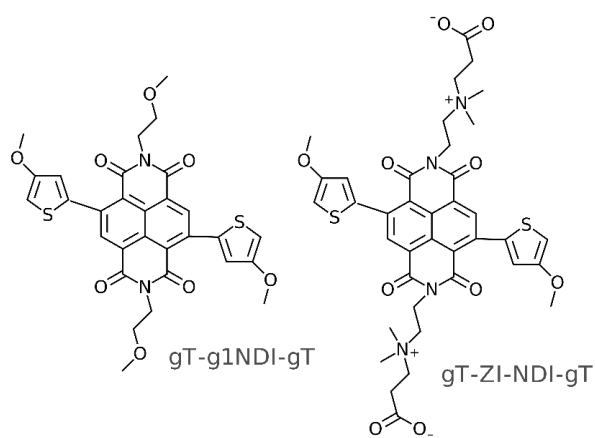

Figure S27. Structural formulas of calculated monomers of gT-g1NDI-gT and gT-ZI-NDI-gT.

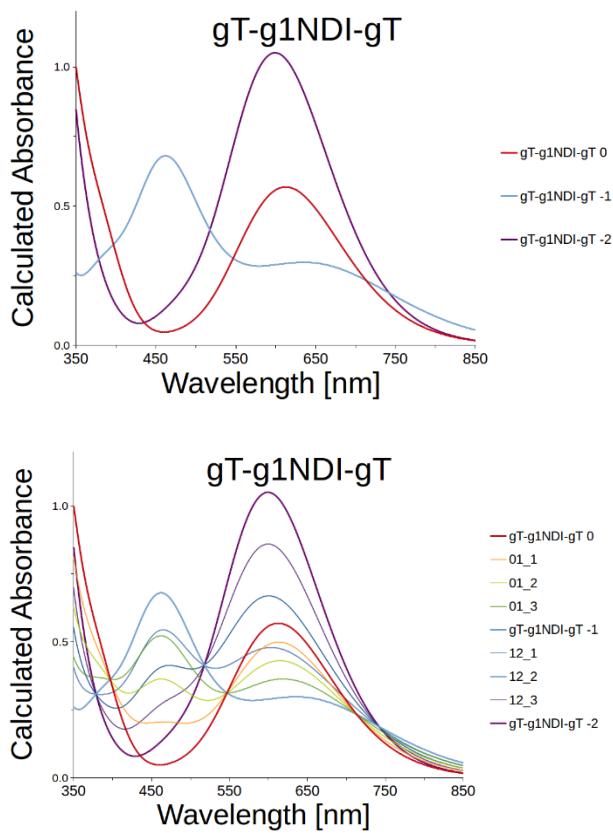

Figure S28. (top) TD-DFT calculated absorption spectra of the monomer gT-g1NDI-gT in its neutral, singly and doubly reduced state normalized to the spectra of neutral molecule. (bottom) Linear combination of the spectra qualitatively reproducing results shown in Supplementary Section 18 for p(g7NDI-gT2). In the legend eg. '01_2' means the second (2) out of three step between neutral (0) and singly reduced (1) molecules. Other names were created analogously.

Table S2: Energies and oscillator strengths calculated for gT-g1NDI-gT, gT-g1NDI-gT$^{-1}$ and gT-g1NDI-gT$^{-2}$.

| **Neutral gT-g1NDI-gT** | | gT-g1NDI-gT$^{-1}$ | | gT-g1NDI-gT$^{-2}$ | |
|---|---|---|---|---|---|
| Energy [eV (nm)] | Oscillator strength | Energy [eV (nm)] | Oscillator strength | Energy [eV (nm)] | Oscillator strength |
| 2.02 (613.7) | 0.25 | 1.80 (687.0) | 0.09 | 1.98 (627.2) | 0.01 |
| 2.21 (562.3) | 0.02 | 2.07 (598.2) | 0.06 | 2.07 (600.3) | 0.45 |
| 3.14 (394.7) | 0.03 | 2.36 (524.6) | 0.05 | 2.55 (486.7) | 0.06 |
| | | 2.58 (480.3) | 0.02 | 3.02 (410.6) | 0.01 |
| | | 2.68 (462.2) | 0.24 | | |
| | | 2.88 (430.4) | 0.02 | | |

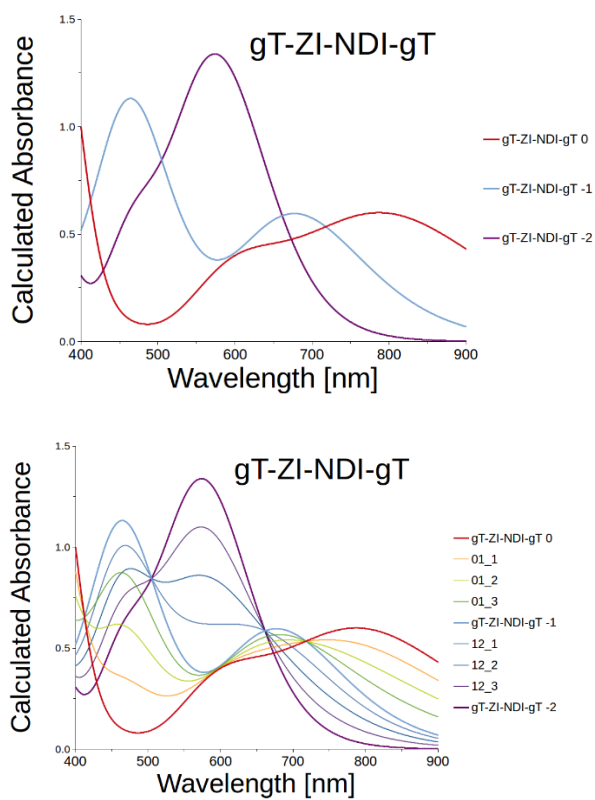

Figure S29: (top) TD-DFT calculated absorption spectra of the monomer gT-ZI-NDI-gT in its neutral, singly and doubly reduced state normalized to the spectra of neutral molecule. (bottom) Linear combination of the spectra qualitatively reproducing results shown in the main text for p(ZI-NDI-gT2). In the legend eg. '01_2' means the second (2) out of three step between neutral (0) and singly reduced (1) molecules. Other names were created analogously

Table S3: Energies and oscillator strengths calculated for gT-ZI-NDI-gT, gT-ZI-NDI-gT$^{-1}$ and gT-ZI-NDI-gT$^{-2}$.

| Neutral gT-ZI-NDI-gT | | gT-ZI-NDI-gT $^{-1}$ | | gT-ZI-NDI-gT $^{-2}$ | |
|---|---|---|---|---|---|
| Energy [eV (nm)] | Oscillator strength | Energy [eV (nm)] | Oscillator strength | Energy [eV (nm)] | Oscillator strength |
| 1.47 (843.4) | 0.05 | 1.70 (728.8) | 0.01 | 2.14 (579.2) | 0.35 |
| 1.57 (789.7) | 0.11 | 1.83 (679.2) | 0.16 | 2.21 (560.2) | 0.01 |
| 2.02 (613.4) | 0.09 | 2.38 (522.0) | 0.04 | 2.63 (471.9) | 0.16 |
| 2.07 (596.8) | 0.01 | 2.39 (518.8) | 0.01 | | |
| 2.27 (546.3) | 0.01 | 2.48 (500.5) | 0.03 | | |
| 2.69 (460.5) | 0.02 | 2.61 (474.9) | 0.03 | | |
| | | 2.66 (465.3) | 0.19 | | |
| | | 2.85 (435.0) | 0.05 | | |
| | | 2.93 (423.1) | 0.01 | | |
| | | 3.01 (411.4) | 0.05 | | |

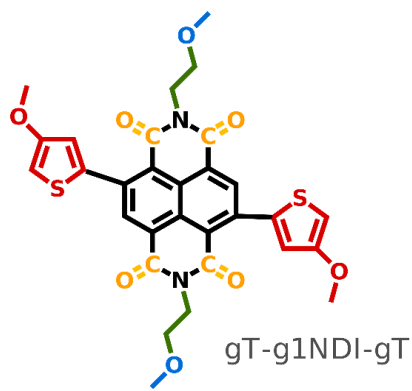

Figure S30. Molecules' sections of gT-g1NDI-gT proposed to clarify charge distribution presented in Table S4.

Table S4. Calculated charge distribution on selected fragments of gT-g1NDI-gT (colored in Figure S30) in neutral, polaron and bipolaron states (columns 2, 3 and 5). Columns 4 and 6 consist of differential charge, being the difference in charge distribution between neutral and polaron states and bipolaron and polaron states respectively.

| Molecule's fragment | | gT-g1NDI-gT Neutral | gT-g1NDI-gT Polaron -1 | Charge difference between **polaron** and neutral state | gT-g1NDI-gT Bipolaron -2 | Charge difference between **bipolaron** and polaron state |
|---|---|---|---|---|---|---|
| (CH$_2$O)$_2$ | | -0.4 | -0.47 | -0.07 | -0.55 | -0.07 |
| (C$_2$H$_4$)$_2$ | | +0.73 | +0.74 | -0.01 | +0.73 | -0.00 |
| C=O | C$_4$ | +2.32 | +2.27 | -0.09 | +2.07 | -0.16 |
| | O$_4$ | -1.77 | -1.99 | -0.22 | -2.22 | -0.23 |
| NDI (without C=O) | | -0.84 | -1.21 | -0.37 | -1.49 | -0.28 |
| (gT)$_2$ | | -0.03 | -0.29 | -0.25 | -0.54 | -0.25 |

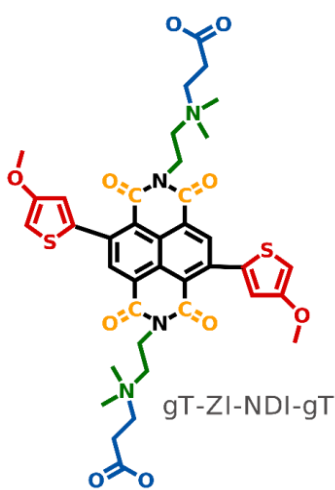

Figure S31. Molecules' sections of gT-ZI-NDI-gT proposed to clarify charge distribution presented in Table S5.

Table S5. Calculated charge distribution on selected fragments of gT-ZI-NDI-gT (colored in Figure S31) in neutral, polaron and bipolaron states (columns 2, 3 and 5). Columns 4 and 6 consist of differential charge that is the difference in charge distribution between neutral and polaron states and bipolaron and polaron states respectively.

| Molecule's fragment | | gT-ZI-NDI-gT Neutral | gT-ZI-NDI-gT Polaron -1 | Charge difference between **polaron** and neutral state | gT-ZI-NDI-gT Bipolaron -2 | Charge difference between **bipolaron** and polaron state |
|---|---|---|---|---|---|---|
| $(C_3H_4O_2)_2$ | | -1.13 | -1.34 | -0.21 | -1.44 | -0.1 |
| $(C_4H_{10}N)_2$ | | +1.51 | +1.67 | +0.16 | +1.52 | -0.15 |
| C=O | $C_4$ | +2.52 | +2.45 | -0.18 | +2.08 | -0.37 |
| | $O_4$ | -1.89 | -2.1 | -0.21 | -2.27 | -0.17 |
| NDI (without C=O) | | -0.98 | -1.44 | -0.46 | -1.48 | -0.04 |
| $(gT)_2$ | | -0.02 | -0.23 | -0.21 | -0.41 | -0.17 |

# 10. Reversibility and electrochemical charging of p(gT2) p-type polymer

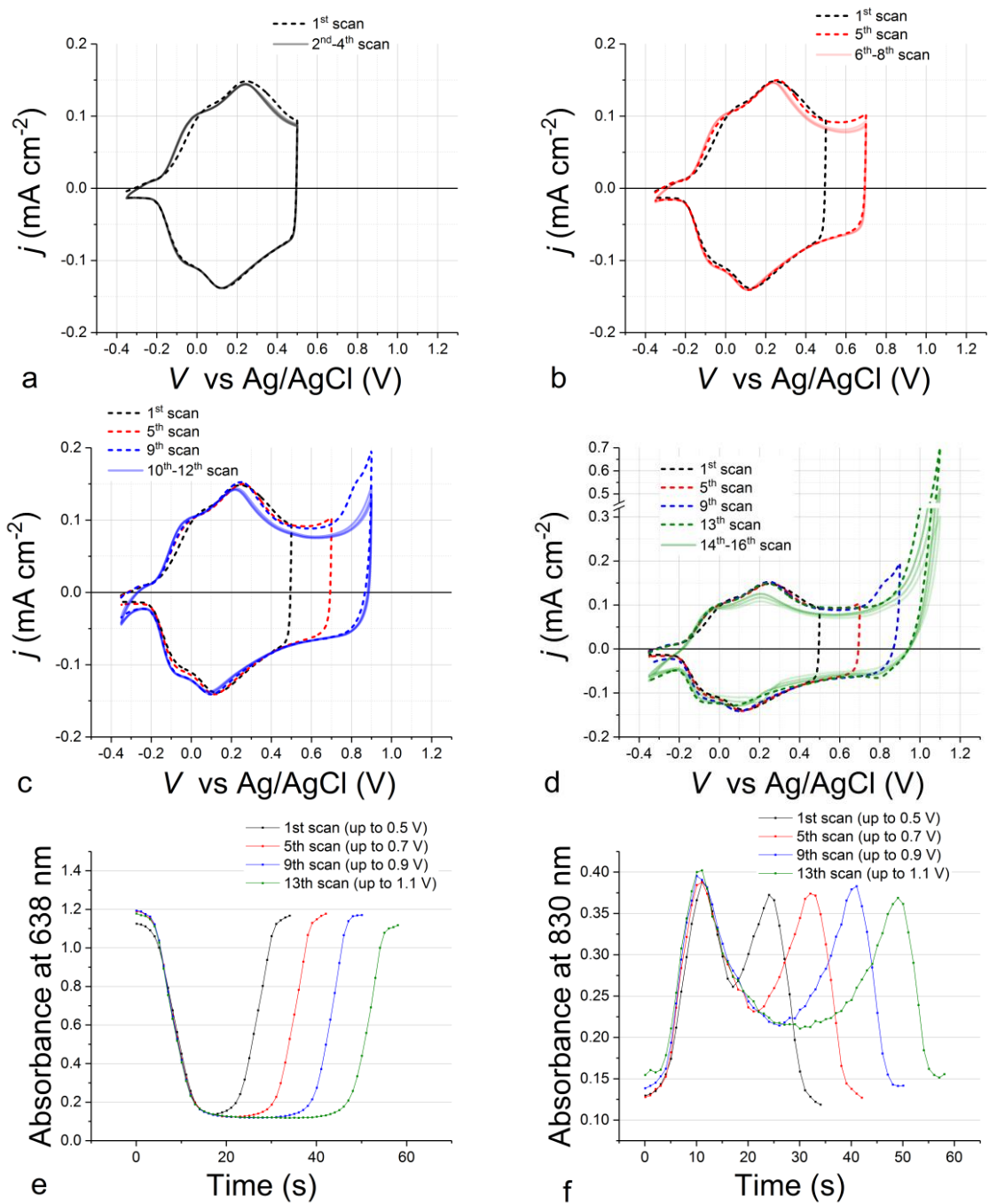

Figure S32. (a) to (d) consecutive cyclic voltammetry measurements performed on a p(gT2) polymer film deposited on FTO in 0.1 M NaCl:DIW at scan rate of 50 mV s$^{-1}$. Absorbance evolution monitored at (e) 638 nm and (f) 830 nm as a function of time during the CV scans shown in (a)-(d).

Figure S32 (a)-(d) shows cyclic voltammetry measurements of p(gT2) for different maximum positive potential reached during the charging of the film. Reversibility of the scans remains good up to applied potentials of 0.9 V vs Ag/AgCl. In Figure S32 (e) and (f) the absorbance at 638 nm and 830 nm for scans up to different potentials are shown. We observe that formation of bipolarons in the film which can be monitored as the drop in absorbance at 830 nm after the first peak in Figure S32 (f), appears to complete at voltages of about 1.1 V.

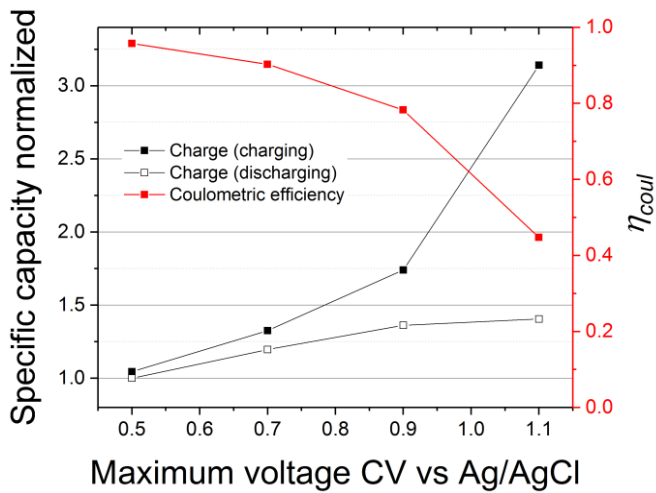

Figure S33. Specific capacity and coulombic efficiency of the same p(gT2) film scanned up to different maximum potential vs Ag/AgCl of a cyclic voltammetry measurement.

Figure S33 shows that when scanning the p(gT2) film at different maximum positive potentials (as shown above in Figure S32), the calculated specific capacity increases. Specifically, the specific capacity calculated from the charge extracted during the discharging cycle for the scan up to 1.1 V vs Ag/AgCl is about 1.4 times larger than the one extracted when scanning up to 0.5 V vs Ag/AgCl. However, the coulombic efficiency undergoes a significant drop when increasing the maximum voltage going from ~1 for measurements scanning up to 0.5 V vs Ag/AgCl to <0.5 when scanning up to 1.1 V vs Ag/AgCl.

# 11. Reversibility of p(ZI-NDI-gT2) n-type polymer

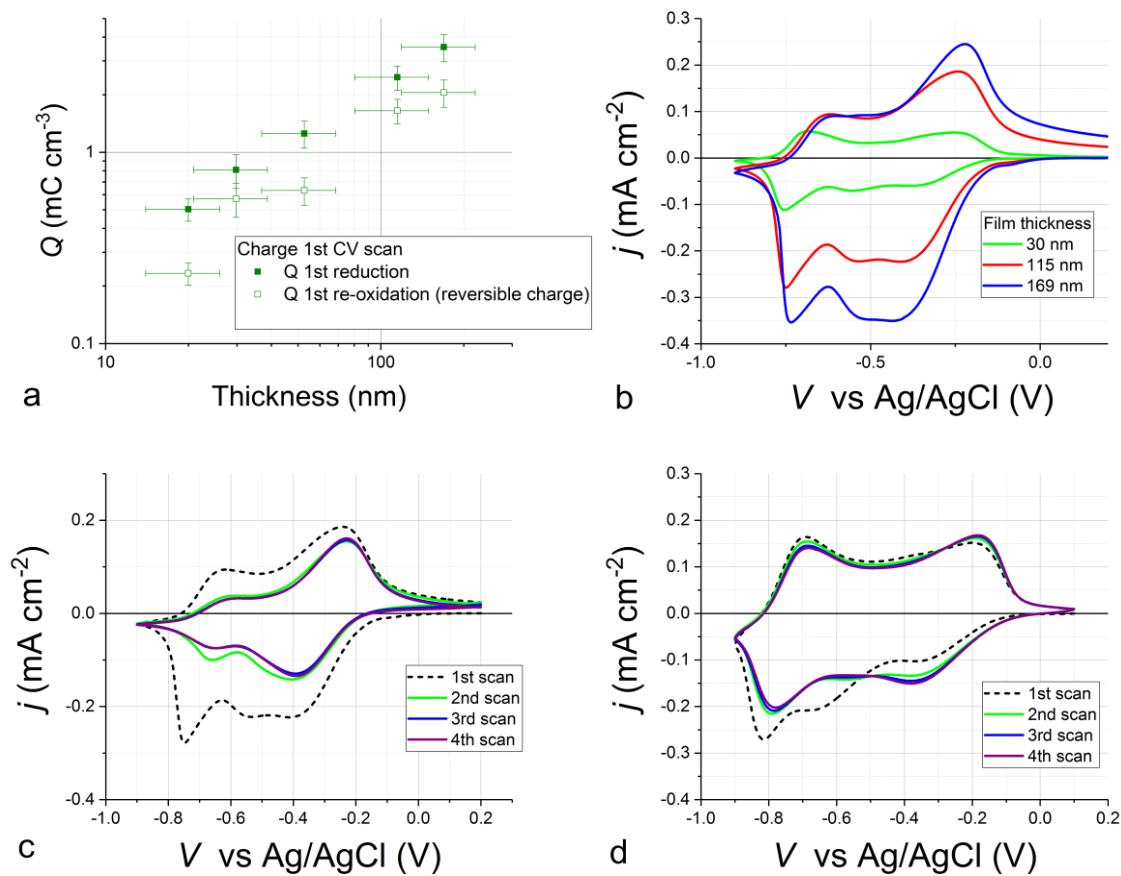

Figure S34. (a) Integrated charge of reduction and reversible re-oxidation of p(ZI-NDI-gT2) deposited on FTO from a methanol solution. Data in (a) are displayed as a function of film thickness and correspond to the 1st cyclic voltammetry measurements on the films (examples are plotted in (b)). (c) and (d) show comparison between the first 4 cyclic voltammetry for p(ZI-NDI-gT2) deposited on FTO from (c) methanol and (d) DMSO.

For the thickness dependent characterization of p(ZI-NDI-gT2) we used a methanol based solution which enabled the deposition of more uniform films compared to DMSO. On the other hand the reversibility of the films was found to be different for the two solvents, possibly due to differences in film morphology (see Figure S34c and d). Figure S34a shows the total charge injected and extracted from films deposited from methanol during the first cyclic voltammetry measurement, showing a significant difference between the two values, also evident from Figure S34b. Films deposited from DMSO were used in for the characterization shown in the main text.

# 12. Thickness dependent specific capacity

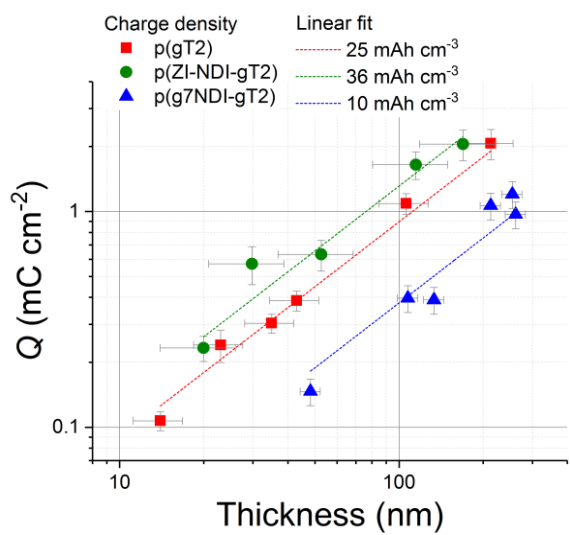

Figure S35. Charge density as a function of film thickness. The dotted lines are linear fits with slope 1 on a log-log scale to the data.

Figure S35 shows the thickness dependent charge density of the three polymers presented in this study and extracted from cyclic voltammetry performed at 50 mV s$^{-1}$. For the p(gT2) and the p(g7NDI-gT2) polymers, the reversible charge of the second CV was taken for their respective datasets. For p(ZI-NDI-gT2) the reversible charge from the 1$^{st}$ CV was instead considered, probably representing an underestimate of the total charge that can be accumulated in films deposited from DMSO as discussed in the previous section. The three datasets show a close to linear relation between charge density and thickness in the range 10 to 200 nm (the fits are run with slope constraint to 1), suggesting that charging and discharging occurs for all the polymers in the second timescale. The resulting specific densities are shown in the legend. These values are conservative estimates of the specific capacity for the polymers for the following reasons:

- For the p(gT2) polymer a value of 25 mAh cm$^{-3}$ was found considering scanning of the polymer up to 0.5 V vs Ag/AgCl. Based on the discussion above we can conclude that this is an underestimate of the potential specific capacity of p(gT2) polymer films in that larger values are expected when scanning to more positive potentials (up to 35 mAh cm$^{-3}$ when scanning up to 1.1 V vs Ag/AgCl). One expects however the coulombic efficiency to drop significantly under these conditions as shown in Figure S32.
- For p(g7NDI-gT2) we consider only the charge accumulated in films scanned to -0.55V vs Ag/AgCl.
- For p(ZI-NDI-gT2) we showed in the previous section that films deposited from methanol show lower reversibility during the first CV scan compared to those deposited from DMSO. The charge injected in the film during the first cyclic voltammetry measurement (displayed in Figure S34) would yield specific capacity in the order of 65 mAh cm$^{-3}$.

The calculation of the number of charges per monomer unit for each polymer in the charged state in the main text was performed by dividing the electronic charge density (extracted from the above measurements) by the density of monomers in the film. This was done assuming a density of the polymers of 1 g cm$^{-3}$.

# 13. Stability measurements

The stability of polymer films deposited on FTO was measured using cyclic voltammetry at 50 mV s$^{-1}$ scan rate. In Figure S36 we show some of the CV scans. The p(gT2) and the p(ZI-NDI-gT2) polymers show qualitatively unchanged electrochemical behavior over 1000 cycles with only a <30% drop in the reversible charge exchanged with the electrode during the experiment (compared to the 2$^{nd}$ scan). The p(g7NDI-gT2) polymer shows similarly good stability when scanned to -0.55 V vs Ag/AgCl. When the polymer is scanned to more negative potentials, the shape of the CV undergoes significant changes. First, the second peak observed at about -0.6 V vs Ag/AgCl during the first scan is not present in any of the following scans, as discussed in the previous sections. Secondly, the negative peak at about -0.5 V vs Ag/AgCl and the positive peak at about -0.35 V vs Ag/AgCl shift apart. The total charge exchanged with the electrode decreases more significantly in this case with a drop of 30% after 50 cycles (compared to the 2$^{nd}$ scan). This drop is not necessarily related to degradation of the polymer in terms of active redox sites. The change in shape of the CVs suggest that transport of electronic or ionic charge might be compromised or that changes in the interfacial kinetics of transfer occur over cycling.

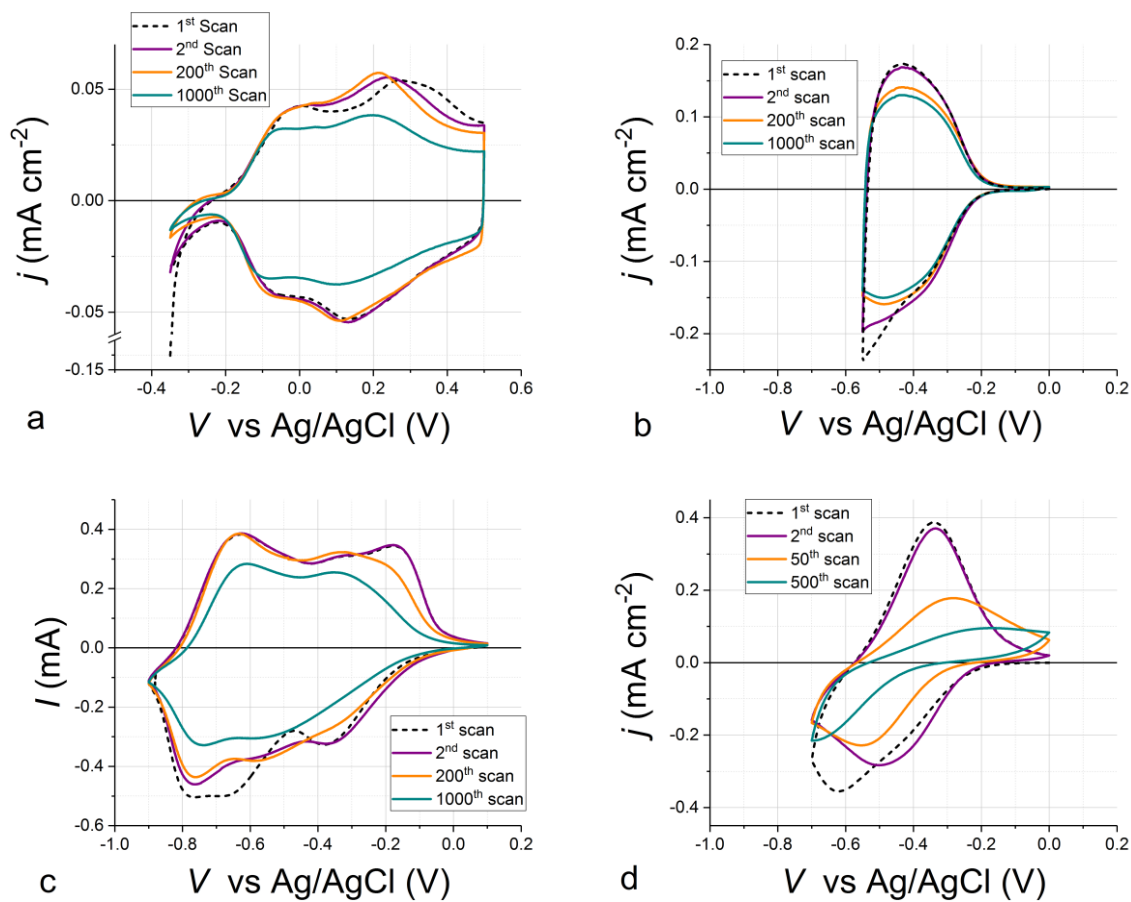

Figure S36. Stability of (a) p(gT2) (b) p(g7NDI-gT2) and (c) p(ZI-NDI-gT2) polymers upon continuous cycling.

# 14. Current peak vs scan rate for p(gT2)/p(ZI-NDI-gT2) device

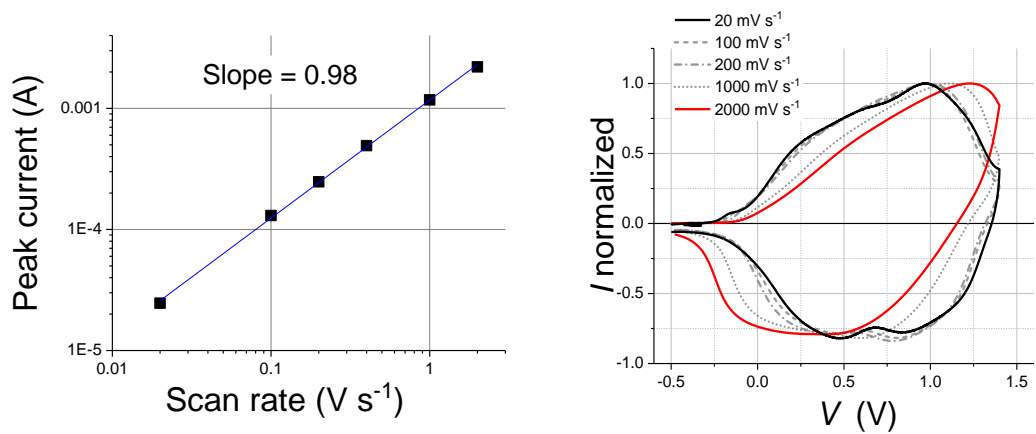

Figure S37. (left) Current peak vs scan rate for the battery device presented in the main text. (right) CV measurements performed at scan rates between 20 and 2000 mV s$^{-1}$

# 15. Rate capabilities

Figure S38 shows the dynamic response of the polymer electrodes using galvanostatic charging-discharging cycling.

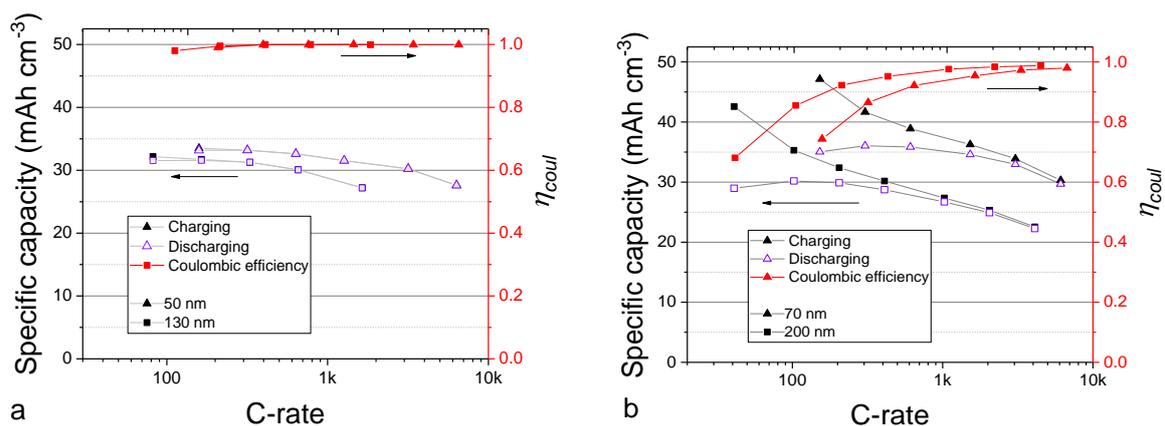

Figure S38. Charge density and coulombic efficiency for two (a) p(gT2) and (b) p(ZI-NDI-gT2) films with different thicknesses extracted from galvanostatic charging discharging measurements displayed as a function of C-rate.

The limited performance in terms of coulombic efficiency at slow C-rates of the p(ZI-NDI-gT2) polymer films are attributed to the presence of oxygen in the electrochemical cell used for three electrode measurements in our setup.

# 16. Impedance measurements

Figure S39a, c and b, d show the impedance measurements run at three different electrochemical potentials corresponding to different charge states of a 40 nm thick p(gT2) film and 70 nm thick p(ZI-NDI-gT2) films respectively on FTO.

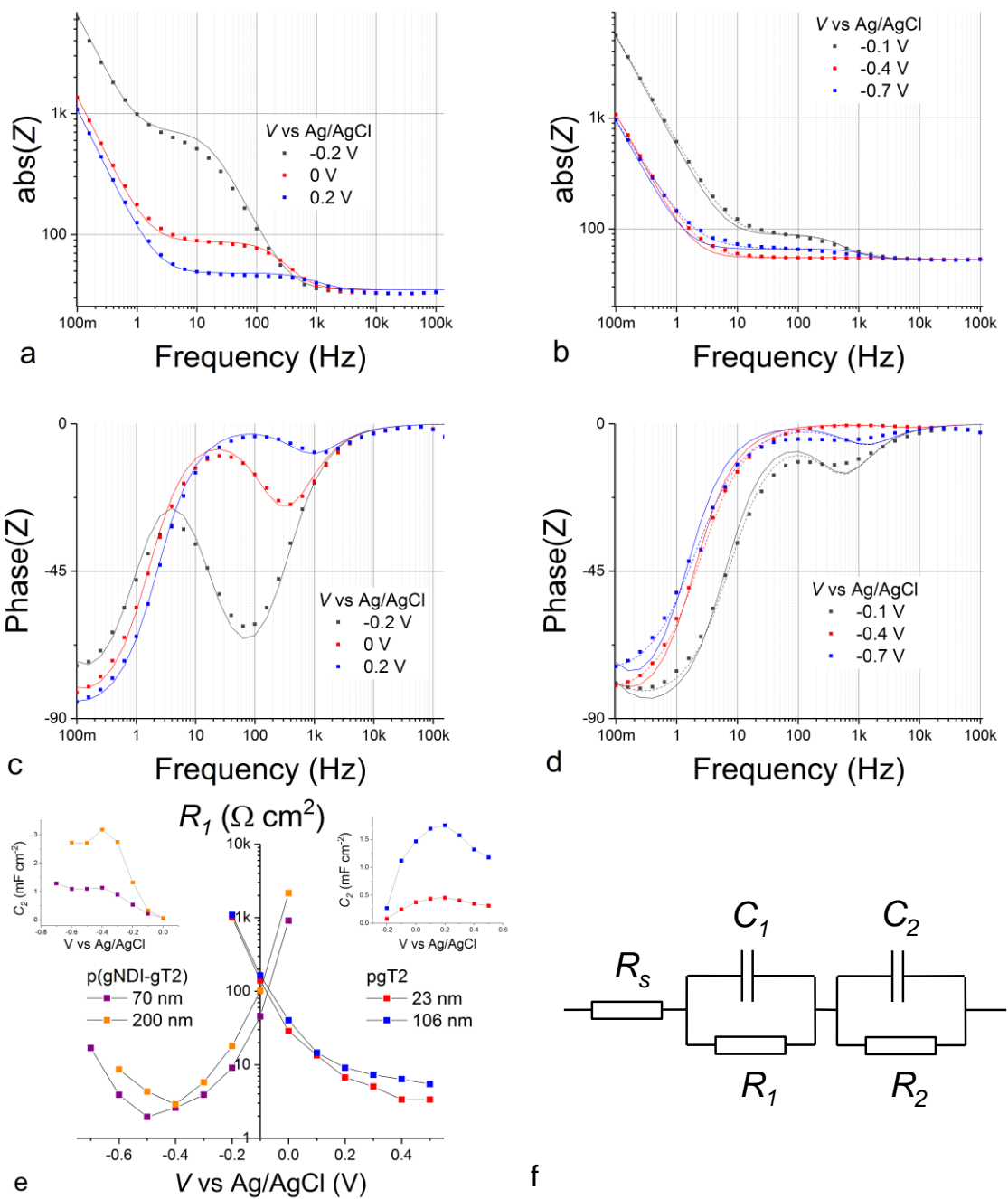

Figure S39. Electrochemical impedance spectroscopy measurements on (a) (c) p(gT2) and (b) (d) p(ZI-NDI-gT2) polymers deposited on FTO in 0.1 M NaCl:DIW at different potentials. (e) Plot of the transfer resistance $R_1$ as a function of applied potential for different thicknesses of the p(gT2) or p(ZI-NDI-gT2) layer. Inset of (e) shows the differential capacitance $C_2$ which is attributed to the electrochemical capacitance of the electrode. (d) Equivalent circuit fitted to the impedance data. For each sample, the series resistance $R_1$ and the interfacial capacitance $C_1$ were kept fixed in the fitting to values in the order of $R_1$ = 50 Ω and $C_1$ = 5 - 10 μF cm$^{-2}$ for measurements at different potentials.

The equivalent circuit used for the fit is illustrated in Figure S39f and includes a series resistance ($R_s$), an interfacial RC parallel circuit ($R_1$, $C_1$), an electrochemical capacitor ($C_2$) in parallel to a leakage resistor to describe retention limitations ($R_2$). The charge transfer resistance $R_1$ decreases significantly with increasing electrochemical doping level for both polymers (increasing potential for p(gT2) and decreasing potential for p(ZI-NDI-gT2)) as shown in Figure S39e. Increased doping in the organic semiconductor is expected to decrease the contact resistance for a metal semiconductor interface in the solid state.[5] In addition, gradual accumulation of ionic charge at the FTO/polymer interface might facilitate the electronic charge injection. Changes in the ion penetration rate at the polymer/electrolyte interface would also give rise to similar behavior and cannot be ruled out. The capacitance $C_2$ represents the electrochemical capacitance of the polymer. Its dependence on the applied potential and its magnitude (see inset in Figure S39e) is consistent with the shape of the cyclic voltammetry measurement and the thickness dependent capacity of the polymers. The series resistance ($R_s$) and interfacial capacitance ($C_1$) were kept constant in the global fit to values of about 50 Ω, consistent with the expected FTO resistance, and 10 μF cm$^{-2}$, which is in the order of a double layer capacitance. This also supports the choice of the equivalent circuit used for the analysis. While the circuit in Figure S39f can well describe the behavior of p(gT2), we find that for p(ZI-NDI-gT2), substituting $C_2$ with a constant phase element yields slightly better fits (dashed lines in Figure S39d). This is consistent with transport limitation for this polymer which would reflect in deviation from a perfectly capacitive behavior (neglecting the contribution of $R_2$) in the low frequency region of the impedance spectrum. Such deviation becomes more pronounced at potentials corresponding to the second reduction peak of the polymer, suggesting that the transport properties of the film are affected by high electron and ion density (for this reason the plot of $C_2$ for p(ZI-NDI-gT2) is interrupted at voltages of -0.6 and -0.7 V vs Ag/AgCl for the two samples shown in Figure S39e).

# 17. Spectrochronocoulometry measurements

Figure S40 shows spectrochronocoulometry measurements associated to step potential between -0.3 V and 0 V vs Ag/AgCl for p(gT2) and between 0 V and -0.5 V vs Ag/AgCl for p(ZI-NDI-gT2). Figure S40 a and b show that the integrated charge changes rapidly in the seconds timescale and show further variation for t > 1 s. This slow component could be related to slow charging of the electrode or to charges that are lost in the electrolyte due to charge retention limitations. Monitoring the optical signal in the millisecond timescale enables an accurate analysis of the polymer charging dynamics, in that only charges that are injected into the polymer are visible at wavelengths corresponding to polaron absorption. The transient electroabsorption measurements referring to the chronoamperometry traces in Figure S40a and b are displayed in Figure S40 c and d. Monoexponential functions fitted well most of the transient absorption profiles and showed time constants in the order of 0.1 s to 1 s for films of about 15 to 100 nm thickness. Also, no slow tail in the absorption trace is detected, suggesting that the slow charging observed from the electrical measurements is related to retention limitations or possibly a slow ionic uptake by the polymer film.

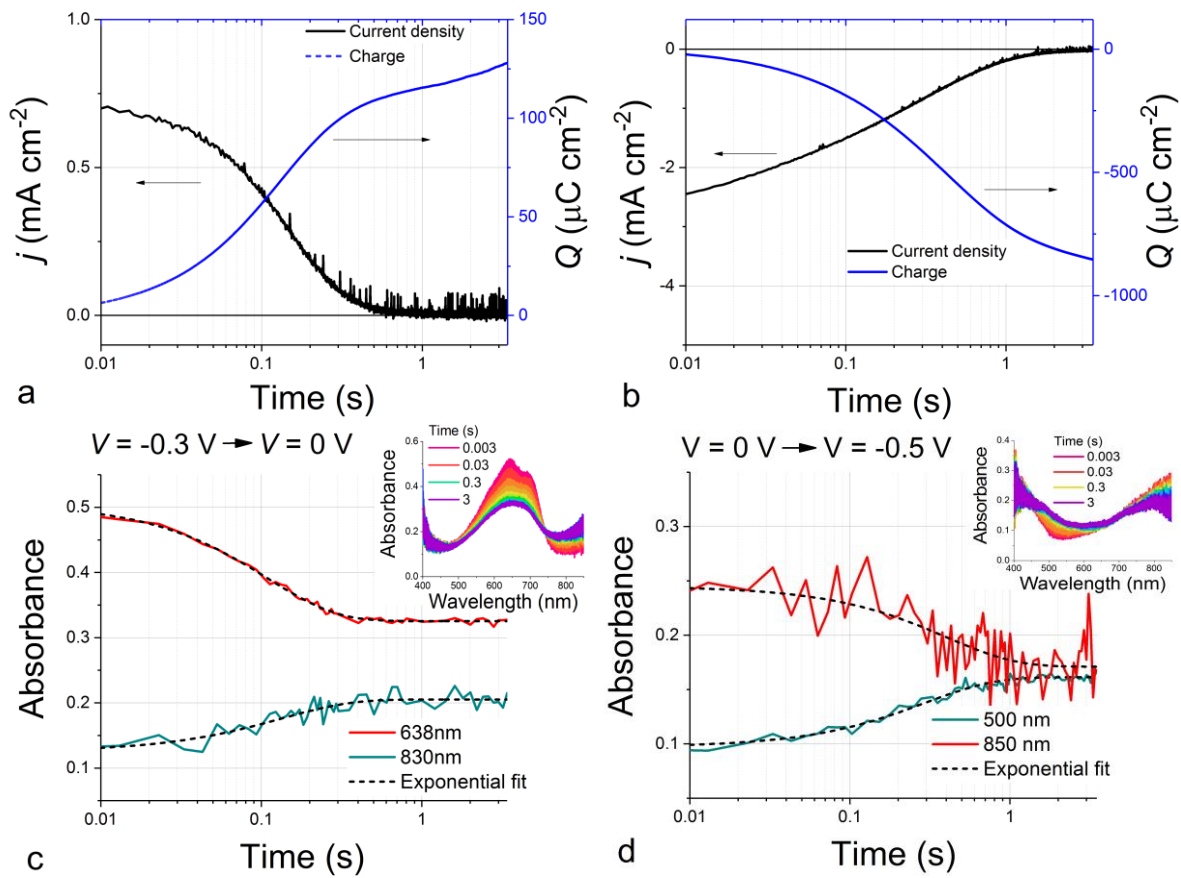

Figure S40. Spectrochronocoulometry of (a) (c) p(gT2) and (b) (d) p(ZI-NDI-gT2) polymers on FTO performed in 0.1 M NaCl:DIW.

The thickness dependence of the time constant extracted from fitting monoexponential functions to the time resolved absorption in spectrochronocoulometry measurements for p(gT2) is shown in Figure S41.

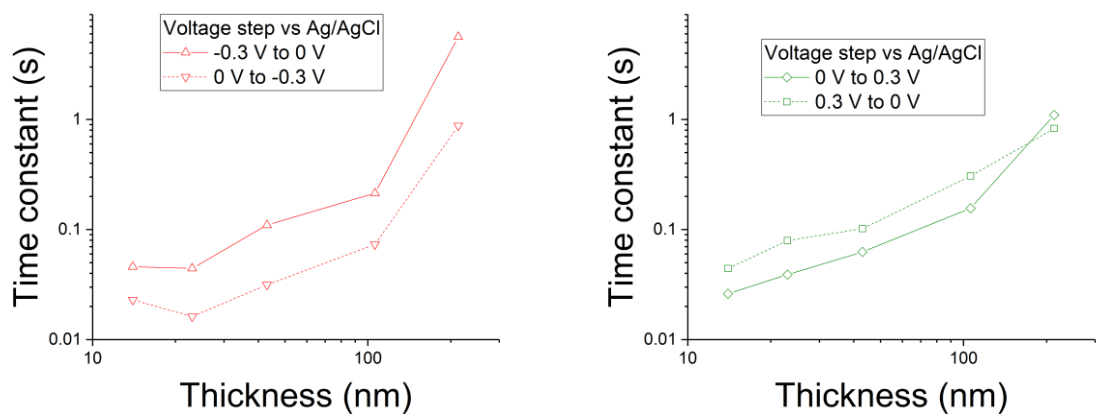

Figure S41. Time constant calculated for charging and discharging spectrochronocoulometry experiments performed on p(gT2) films of different thickness. (a) First peak (step potential -0.3 V to 0 V vs Ag/AgCl); (b) second peak (step potential 0 V to 0.3 V vs Ag/AgCl).

The data show that for all films fabricated up to about 100 nm, linear dependence of the fitted time constant on thickness is detected. This observation rules out diffusion limitation in the charging/discharging mechanism, in that such regime would manifest itself in a square dependence of the time constant on thickness. This suggests that electronic and ionic transport in the film is not a limiting factor for the maximum charging/discharging rate achievable with this polymer, consistent with the discussion of impedance measurements presented in the previous section.

# 18. Reversibility and electrochemical charging of p(g7NDI-gT2) n-type polymer

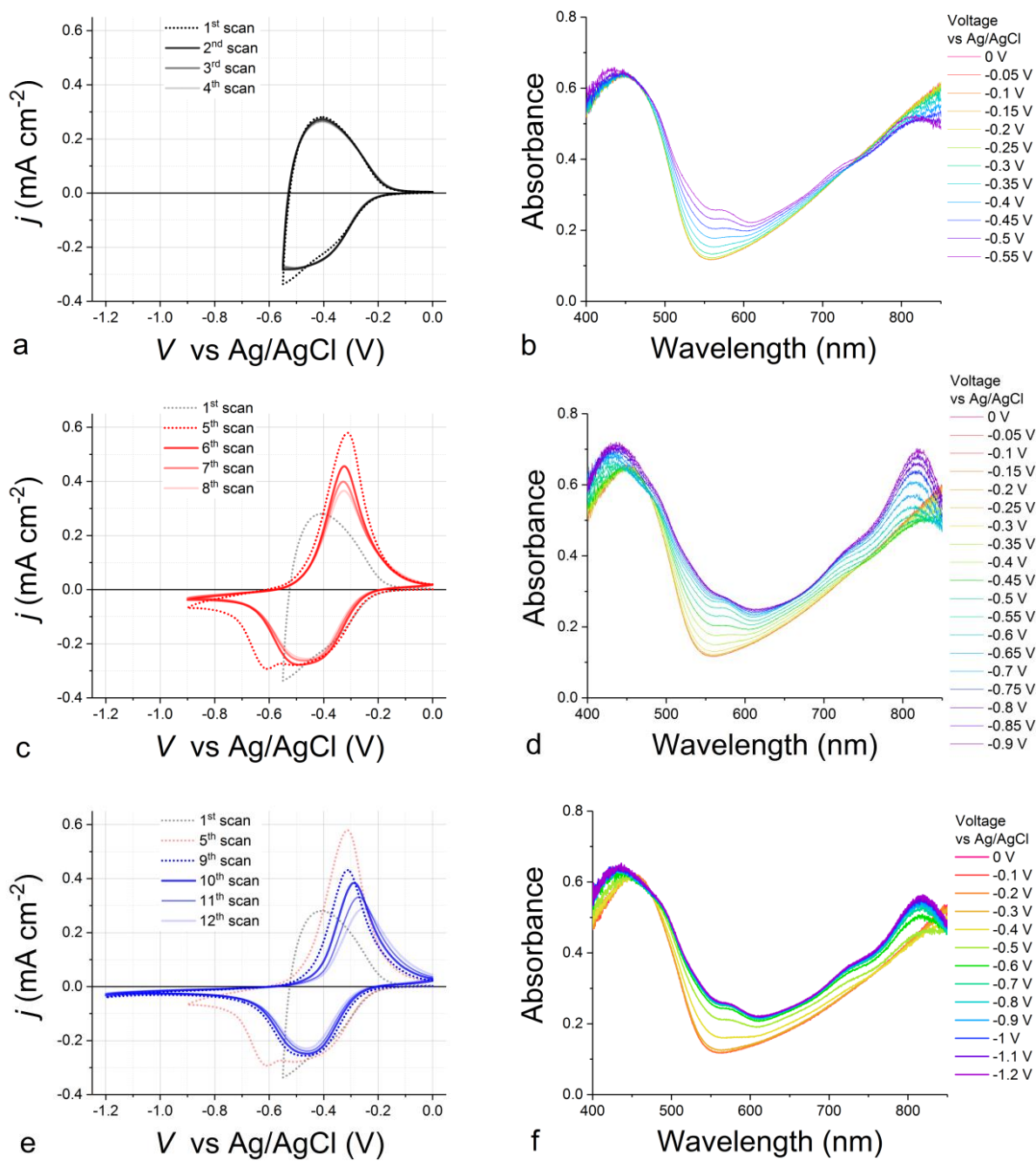

Figure S42. (a),(c),(e) consecutive cyclic voltammetry measurements performed on a p(g7NDI-gT2) polymer film deposited on FTO in 0.1 M NaCl:DIW at scan rate of 50 mV s$^{-1}$ reaching a negative potential of (a) -0.55 V, (c) -0.9 V, (e) -1.2 V vs Ag/AgCl. Film absorbance evolution referred to the (b) 1$^{st}$, (d) 5$^{th}$ and (f) 9$^{th}$ scan.

Figure S42 shows spectroelectrochemical measurements of a p(g7NDI-gT2) film scanned within different potential ranges. The presence of a second peak when scanning to potential more negative than -0.55 V vs Ag/AgCl is shown in Figure S42c. This peak is not there in the following scans as discussed in the main text. Also, the changes in absorbance detected for the first measurement scanning down to -0.9 V vs Ag/AgCl (5$^{th}$ scan in Figure S42c) are greater than the ones appearing over the following cycles. This suggests that the ability of the film to reversibly reach deep level of charging changes after the first scan to potentials more negative than -0.55 V vs Ag/AgCl. The spectral signatures that we observe at potentials more negative than about -0.4 V vs Ag/AgCl could be ascribed to bipolaron bands.

# 19.   Charge retention and implications on stability

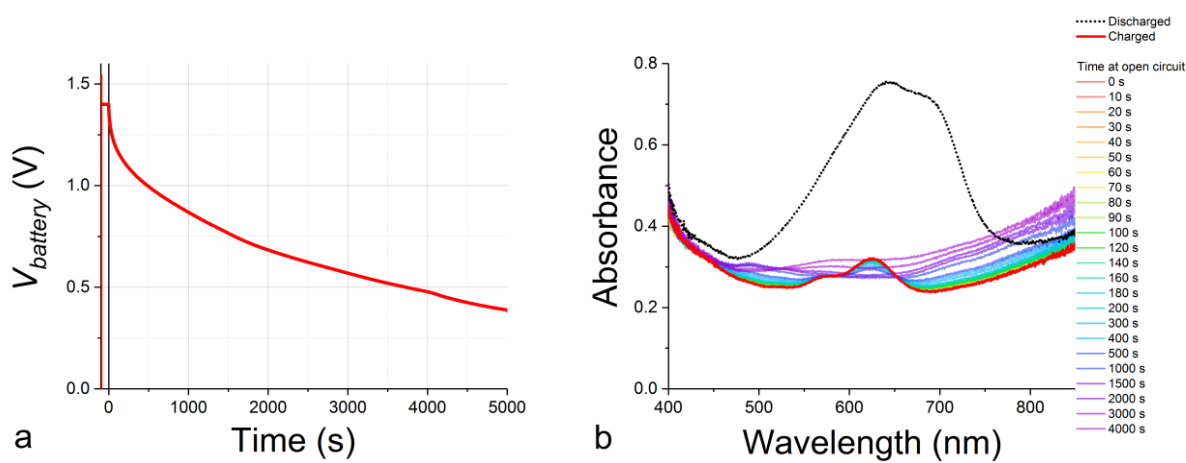

Figure S43. Charge retention experiment performed on a (p(gT2) / 0.1 M NaCl:DIW / p(ZI-NDI-gT2)) battery device. (a) Voltage of the battery as a function of time: the voltage of the device was first kept at 1.4 V for 100 s (Time < 0 s); at Time = 0 s the device was switched to open circuit and its voltage was measured. (b) Optical absorbance of the two polymer films in series as a function of time.

Figure S43 shows a charge retention experiment run on the battery device presented in the main text. After keeping the voltage at 1.4 V for a 100 s period, the device's open circuit voltage and optical absorbance was monitored as a function of time. A fast voltage drop is observed in the first ~200 s of the experiment, followed by a much slower decay over thousands of seconds. From the spectral analysis we can conclude that the 2$^{nd}$ polaron on the p(ZI-NDI-gT2) polymer is lost first, while it appears that the second polarons on the p(gT2) are longer lived. This is consistent with loss of electrons in the n-type polymer to residual oxygen in the cell. As a result, we expect the polymer battery to potentially suffer from charge unbalance during its operation, in that the charge density in the two polymer films is lost at different rate. This implies that the absolute values of the electrochemical potentials in the two films for a certain voltage is function of the battery history. This effect is visible when performing continuous scanning of the device, as shown in Figure S44, where the features observed in cyclic voltammetry measurements shift in potential. The optical spectra shown in Figure S44 suggests that the two polymers are in different charged stated at the beginning of each scan. This is consistent to the observation above: as a result of the loss of charges in the n-type polymer, the p-type polymer accumulates holes which cannot leave the film since there are not enough electrons on the anode of the cell. This also implies that the capacity of the device decreases during cycling. This analysis shows that the observed drop in capacity is not due solely to degradation of the polymer electrodes.

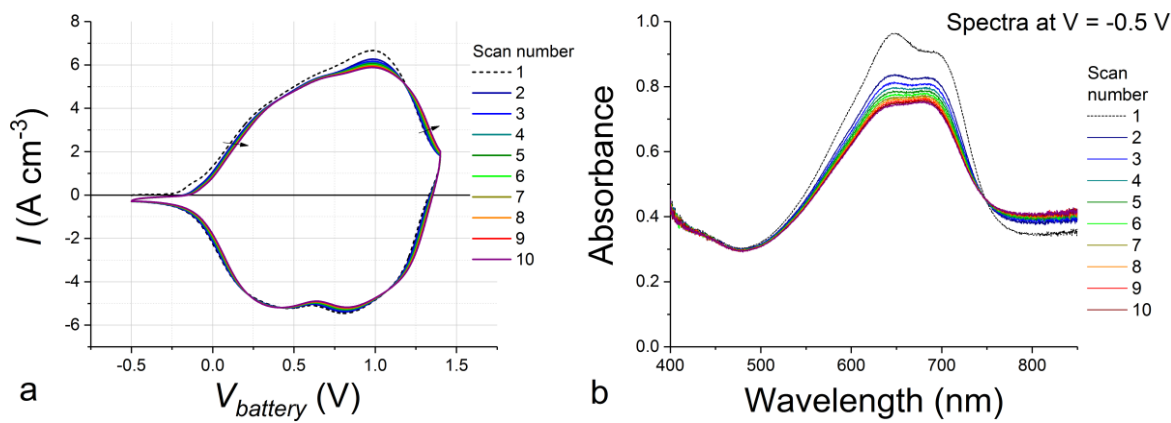

Figure S44. (a) Cyclic voltammetry measurements performed on the (p(gT2) / 0.1 M NaCl:DIW / p(ZI-NDI-gT2)) battery device. (b) Optical absorbance of the battery device measured at the beginning of each scan.

As further prove of this, Figure S45 shows that it is actually possible to attain recovery in capacity of the battery by scanning to negative potentials. The data show that by scanning to -0.8 V the shift in the shift in the CV features is limited over 100 cycles. Moreover, by performing a scan to -1 V, an irreversible peak is observed which we ascribed to the reduction of the p(gT2) chains where holes tend to accumulate during the cycling because of poorer retention performance at the anode. An oxidation process probably involving the electrolyte must happen at the anode to counterbalance this process. Strikingly, the CV performed after this 'recovery' scan shows similar shape and capacity to the one observed for scan 2 of this experiment. We expect that a similar effect played a role in the galvanostatic cycling measurement shown in Figure 4c in the main text.

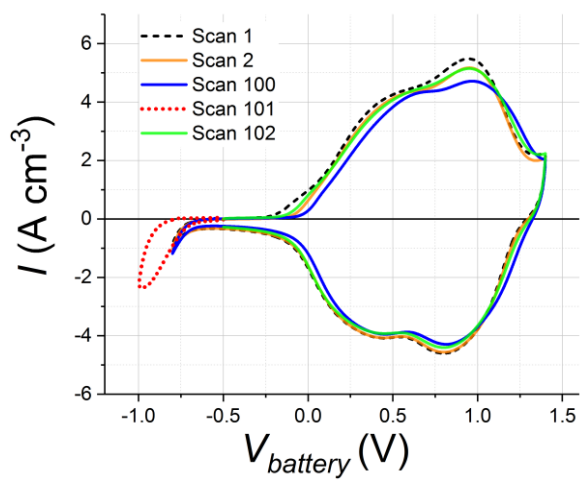

Figure S45. 100 consecutive cyclic voltammetry measurements performed on the (p(gT2) / 0.1 M NaCl:DIW / p(ZI-NDI-gT2)) battery device. The graph shows a recovery effect after scanning the device to more negative potentials.

# Supplementary references